\documentclass[usenatbib,usegraphicx]{mn2e}
\usepackage{aas_macros}
\usepackage{amsmath,amssymb,amsfonts,bm}
\usepackage{natbib}
\usepackage{mathrsfs}
\usepackage{subfigure}
\usepackage{longtable}
\usepackage{graphicx}
\usepackage{graphics}
\usepackage{keyval}
\usepackage{trig}
\usepackage{float}
\usepackage{calc}
\usepackage{url}
\usepackage{hyperref}
\usepackage[dvipsnames]{color}
\def\beq{\begin{equation}}
\def\eeq{\end{equation}}
\def\bey{\begin{eqnarray}}
\def\eey{\end{eqnarray}}
\def\Myr{\, {\rm Myr} }
\def\Gyr{\, {\rm Gyr} }

\def\pc{\, {\rm pc} }
\def\re{\, {\rm R_e} }

\def\kpc{\, {\rm kpc} }
\def\msun{M_\odot}

\def\Lsun{L_\odot}
\def\lsun{L_\odot}
\def\kms{\, {\rm km \, s}^{-1} }
\def\vg{{\bf g}}

\def\dirv{{\bf\nabla\cdot}}

\def\rp{r_{\rm P}}
\def\rh{R_{\rm h}}
\def\rhf{R_{\rm h,f}}
\def\mb{M_{\rm b}}
\def\pdenzero{\Sigma_{\rm 0}}
\def\tcr{t_{\rm cr}}
\def\rt{\mathscr {R}}
\def\lossig{\sigma_{\rm LoS}(R)}
\title[Kinematics of GCs undergoing gas expulsion]{The kinematics of star clusters undergoing gas expulsion in Newtonian and Milgromian dynamics}
\author[Wu \& Kroupa]{Xufen Wu$^{1,2}$\thanks{xufenwu@ustc.edu.cn} and Pavel Kroupa$^{3,4}$\\
  $^1$CAS Key Laboratory for Research in Galaxies and Cosmology, Department of Astronomy, \\
  University of Science and Technology of China, Hefei, 230026, P.R. China\\
  $^2$School of Astronomy and Space Science, University of Science and Technology of China, Hefei 230026, China\\
  $^3$Helmholtz-Institut f\"{u}r Strahlen-und Kernphysik, Universit\"{a}t Bonn, Nussallee 14-16, D-53115 Bonn, Germany\\
  $^4$Charles University in Prague, Faculty of Mathematics and Physics, Astronomical Institute, \\
  V Holesovickach 2, CZ-180 00 Praha 8, Czech Republic
}

\begin{document}

\maketitle

\begin{abstract}
We study the kinematics of stars in clusters undergoing gas expulsion in standard Newtonian dynamics and also in Milgromian dynamics (MOND). Gas expulsion can explain the observed line-of-sight (LoS) velocity dispersion profile of NGC 2419 in Newtonian dynamics. For a given star formation efficiency (SFE), the shapes of the velocity dispersion profiles, which are normalised by the velocity dispersion at the projected half-mass radius, are almost indistinguishable for different SFE models in Newtonian dynamics. The velocity dispersion of a star cluster in the outer halo of a galaxy can indeed have a strong radial anisotropy in Newtonian dynamics after gas expulsion. MOND displays several different properties from Newtonian dynamics. In particular, the slope of the central velocity dispersion profile is less steep in MOND for the same SFE. Moreover, for a given SFE, more massive embedded cluster models result in more rapidly declining central velocity dispersion profiles for the final star clusters, while less massive embedded cluster models lead to flatter velocity dispersion profiles for the final products. The onset of the radial-orbit instability in post-gas-expulsion MOND models is discussed. SFEs as low as a few percent, typical of molecular clouds, lead to surviving ultra-diffuse objects. Gas expulsion alone is unlikely the physical mechanism for the observed velocity dispersion profile of NGC 2419 in MOND.

\end{abstract}


\section {Introduction}
Star clusters form in collapsing cloud clumps within giant molecular clouds \citep[GMCs,][]{Tilley_Pudritz2004}, and most stars are born in embedded star clusters \citep{Lada_Lada2003,Kroupa2005,Kroupa2008,Marks_Kroupa2011}. The timescale of evolution from a pure gaseous protocluster through the embedded cluster phase and to an exposed very young cluster is approximately $1~\Myr$. The gas is expelled from the embedded cluster through multiple mechanisms including stellar winds, type  II supernova explosions, ionizing radiation, radiation pressure and outflows \citep{Hopkins+2013,Dib+2013,Krumholz_Matzner2009,Murray+2010,Hansen+2012}. The residual gas leaves with an outflow velocity of approximately $12\kms$ for the observed Treasure Cheast cluster \citep{Smith+2005} and $25-30\kms$ for the star clusters observed in the Antennae galaxies \citep{Whitmore+1999,Zhang+2001}. An outflow velocity of $10\kms$ is commonly used in simulations \citep{Kroupa+2001,Banerjee_Kroupa2012,Banerjee_Kroupa2015,Banerjee_Kroupa2017,Kroupa+2018}. The star formation stops after gas expulsion. The gravitational potential well of the exposed star cluster is much shallower compared to that of the embedded cluster \citep[e.g.,][]{Goodwin1997,Bastian_Goodwin2006,Goodwin_bastian2006,Boily_Kroupa2003,Geyer_Burkert2001,Baumgardt_Kroupa2007,Marks+2012,Pfalzner_Kaczmarek2013,Brinkmann+2017,Banerjee_Kroupa2017}, and the binding energy for the stars becomes smaller. Hence, the stars in the star cluster move onto elongated orbits. 

NGC 2419, with ${\rm L_{\rm V}}\approx 4.4\times10^5\Lsun$ \citep{Baumgardt+2009}, is one of the brightest and most distant globular clusters (GCs) in the Milky Way \citep{Bellazzini2007,Harris1996}. The Galactocentric distance of NGC 2419 is $89.9~\kpc$. NGC 2419 is a metal-poor GC with a metallicity of $[Fe/H]\approx -2.32\pm 0.11$ dex and an alpha element abundance $[\alpha /Fe]=+0.2$ dex \citep{Shetrone+2001}. The size of NGC 2419, with a projected half-mass radius of $\rh\approx 19\pc$, is about five times larger than that of a normal GC at the same luminosity \citep{Mackey_vandenBergh2005,Ibata+2011a}. Hence, both the background and the internal gravities of NGC 2419 are in the weak field regime, where the gravitational acceleration is much smaller than Milgrom's gravitational constant, $a_0\approx 3.7 {\rm pc/Myr^2}$. In the weak field regime, the gravitation is in the deep Milgromian dynamics \citep[hereafter referred to as MOND,][]{Milgrom1983a} limit, providing a useful platform to test gravitational dynamics.

The velocity dispersion of stars in NGC 2419 has been measured by \citet{Baumgardt+2009} and \citet{Ibata+2011a}. \citet{Baumgardt+2009} found that a model with an isotropic velocity dispersion profile does not agree with the observed velocity dispersion profile in both Newtonian dynamics and Milgromian dynamics. Later, \citet{Sollima_Nipoti2010} found that a radially anisotropic model is required to reproduce the velocity dispersion profile of NGC 2419 in Newtonian dynamics, while the radially anisotropic MOND models do not agree with the observed velocity dispersion profile of \citet{Baumgardt+2009}. Further simulations by \citet{Ibata+2011a,Ibata+2011b} have corroborated that the velocity dispersion of NGC 2419 is extremely radially anisotropic in both Newtonian and Milgromian dynamics. NGC 2419 has also been further extensively studied in both dynamics by other groups \citep{Conroy+2011,Bruens_Kroupa2011,Sanders2012a,Sanders2012b,Derakhshani_Haghi2014}. In general, all the dynamical models are required to be strongly radially anisotropic in the two frameworks. Although an intermediate-mass black hole (IMBH) might also account for the very steep velocity dispersion profile in the central region of a star cluster, the surface density will have a shallow cusp in the center \citep{Baumgardt+2005b}, which is not the case in NGC 2419. The surface density profile of NGC 2419 is flat in the center according to several authors' observations \citep{Trager+1995,Bellazzini2007,Ibata+2011a}.

To explain the velocity dispersion profile of NGC 2419, it is important to explore the possible mechanism for creating the required extremely radial anisotropy in both dynamics in the existing dynamical studies. A possible mechanism to generate such an anisotropy is partial relaxation \citep{Lynden-Bell1967,Bertin_Trenti2003}. A family of models undergoing partial relaxation has been investigated for a sample of GCs including NGC 2419 in Newtonian dynamics \citep{Zocchi+2012}. A system undergoing partial relaxation is strongly radially anisotropic in its outer regions and thus may explain the kinematics of NGC 2419. 

In this work, we shall propose another physically plausible origin of such a strong radial anisotropy. In the process of rapid gas expulsion, due to the weakening of the gravitational potential, the size of a star cluster expands violently and a large number of stars consequently move onto radial orbits \citep{Brinkmann+2017}. The radial orbits generated by gas expulsion might lead to the radial anisotropy of the velocity dispersion. We consider the gas expulsion process in both Newtonian dynamics and MOND, since NGC 2419 is frequently used to test gravity in the two frameworks, as aforementioned.

\begin{table*}
\begin{center}
  \caption{Parameters of ICs for the embedded (pre-gas expulsion) clusters with a Plummer density profile: the $1_{\rm st}$ through $3_{\rm rd}$ columns are the ID of the ICs, the total mass, $M_{\rm b}$, and the Plummer radius, $\rp$. The $4_{\rm th}$ through $6_{\rm th}$ columns are the projected half-mass radius of the ICs ($\rh$) and of the final star clusters after undergoing gas expulsion ($\rhf$). The projected central density, $\pdenzero$ for the ICs and $\Sigma_{\rm 0,f}$ for the final star clusters, are shown in the $7_{\rm th}$ through $9_{\rm th}$ columns, respectively. The $10_{\rm th}$ through $12_{\rm th}$ columns represent the line-of-sight velocity dispersion at projected half-mass radii. The LoS direction is along $x$-axis for the normalisation constants in the $7_{\rm th}-12_{\rm th}$columns. The fraction of the mass of bound stars in the final star clusters to the mass of the initial embedded clusters are listed, respectively, in $13_{\rm th}$ and $14_{\rm th}$ columns for different SFE.}\vskip 0.30cm
\begin{tabular}{lccccccccccccccc}
\hline
ICs& $M_{\rm b}$ & $\rp$ & $\rh$ &$\rhf$&$\rhf$ &  $\pdenzero$&$\Sigma_{\rm 0,f}$ &$\Sigma_{\rm 0,f}$ &  $\sigma_{\rm 0}$ &$\sigma_{\rm 0,f}$ & $\sigma_{\rm 0,f}$& $f_{\rm b}$& $f_{\rm b}$  \\
& $(\msun)$ & $(\pc)$& $(\pc)$ &$(\pc)$  &$(\pc)$ & $(\msun \pc^{-2})$ & $(\msun \pc^{-2})$  & $(\msun \pc^{-2})$ &$(\kms)$ &$(\kms)$&$(\kms)$ & &\\
\hline
$\epsilon$&  & &  &0.5  &0.4 &  & 0.5  & 0.4 & &0.5&0.4 &0.5&0.4 \\
\hline
N1 & $10^7$ &5.0& 5.0 &15.0&28.9 & $8.21\times10^4$ &$6.88\times10^3$ &$1.99\times 10^{3}$ & 27.82 &9.18 &3.75 & 0.73 & 0.34\\
M1 & $10^7$ &5.0& 5.0&16.2 &34.7 & $7.97\times10^4$ &$6.91\times10^3$ &  $1.96\times10^{3}$& 28.08 &11.29 &7.69 & 0.96 & 0.87\\
M2 & $10^6$ &5.0& 5.0 &10.0 &13.3 & $8.21\times10^3$ &$1.18\times10^3$ & $6.35\times10^2$ & 9.28 &5.55 & 4.80 & 1.00 & 0.98\\
M3 & $10^5$ &5.0& 5.0 &7.1 &8.0 & $8.67\times10^2$&$2.16\times10^2$ &$1.25\times10^2$ & 3.66 & 2.77& 2.58 & 0.99 & 0.98\\
M4 & $10^5$ &10.0&10.0&12.6 &13.7 & $2.13\times10^2$&$6.34\times10^1$ & $3.85\times 10^1$& 3.23  &2.63 & 2.48 & 0.97 & 0.94\\
\hline
\end{tabular}
\label{plummer}
\end{center}
\end{table*}

\begin{table*}
\begin{center}
  \caption{The observed values of the velocity dispersion profile of NGC 2419 \citep{Ibata+2011a}. }\vskip 0.30cm
\begin{tabular}{lccccccccccccc}
  \hline
  Sample A & & & & & & & &\\
  \hline
  R (pc)& 8.29 & 22.05 & 32.40& 47.83 & 63.18 & 122.54 \\
  $\sigma_{\rm LoS}~(\kms)$& $6.573^{+1.413}_{-0.575}$ & $3.816^{+1.032}_{-0.467}$ & $2.645^{+0.826}_{-0.378}$ & $2.662^{+0.731}_{-0.454}$ & $0.997^{+0.576}_{-0.549}$ & $0.756^{+0.436}_{-0.318}$ \\ 
  $\sigma_{\rm LoS}/\sigma_{\rm \rh=19pc}$ &$1.55^{+0.33}_{-0.14}$&$0.90^{+0.24}_{-0.11}$& $0.62^{+0.19}_{-0.09}$ &$0.63^{+0.17}_{-0.11}$&$0.23^{+0.14}_{-0.13}$&$0.18^{+0.10}_{-0.07}$\\
  \hline
  Sample A+B &&&&& \\
  \hline
  R (pc)& 8.51 & 22.25 & 32.76 & 48.63 & 63.50 & 126.37\\
  $\sigma_{\rm LoS}~(\kms)$& $7.061^{+1.420}_{-0.698}$ & $3.778^{+1.002}_{-0.698}$ & $3.896^{+1.079}_{-0.504}$ & $3.109^{+0.779}_{-0.486}$ &$1.994^{+0.664}_{-0.651}$ & $1.295^{+0.683}_{-0.381}$\\  
  $\sigma_{\rm LoS}/\sigma_{\rm \rh=19pc}$& $1.46^{+0.29}_{-0.14}$ & $0.78^{+0.21}_{-0.09}$ & $0.81^{+0.22}_{-0.10}$ & $0.64^{+0.16}_{-0.10}$ & $0.41^{+0.14}_{-0.13}$ & $0.27^{+0.14}_{-0.08}$\\
\hline
\end{tabular}
\label{kin2419}
\end{center}
\end{table*}

In Milgromian dynamics, there is a phase transition from (quasi-)Newtonian dynamics to MOND in the process of gas expulsion, and the amount of phantom dark matter predicted by MOND increases with the weakening of gravitation.\footnote{Phantom dark matter is the effective but unreal matter generated from the MONDian gravitational potential by solving the standard Newtonian Poission's equation \citep{Milgrom1986,Wu+2008}.}
 For a given star formation efficiency (SFE, hereafter denoted by $\epsilon$), after the gas expulsion, the gravitational potential well of a star cluster is deeper in MOND than that in Newtonian dynamics, i.e., the variation of the gravitational potential is smaller in MOND. It has been found that the size expansion of a star cluster undergoing gas expulsion is smaller in MOND than that in Newtonian dynamics with the same $\epsilon$ \citep[][]{Wu_Kroupa2017}. Accordingly, the stars lose less potential energy in a MONDian system, and the orbits of these stars are less elongated. As a result, for a given $\epsilon$, the radial anisotropy of the star cluster is smaller in MOND. Moreover, a MONDian embedded cluster could leave a bound remnant with a much lower $\epsilon_{\rm crit}$ than a Newtonian system \citep{Wu_Kroupa2017}. An interesting question then arises as to whether or not the gas expulsion can result in an extremely radial anisotropy for the MOND models of NGC 2419.

In this work, the anisotropy profiles of star clusters undergoing gas expulsion is studied in both Newtonian dynamics and MOND. The paper is organised as follows: the $N$-body initial conditions (ICs) for the embedded clusters are constructed in the two dynamics, and a sudden gas expulsion process is applied to the embedded clusters by instantaneously reducing the masses of the $N$-body particles in Sec. \ref{ics}. The models are evolved for $2\Gyr$ and the anisotropy profiles are compared in the two dynamics in Secs. \ref{gas} and \ref{kin}. Since MOND allows a much lower $\epsilon_{\rm crit}$ to leave a bound object, we shall study the anisotropy profiles of the final bound objects with ultra-low $\epsilon$ (Sec. \ref{ultralow}), which are impossible to leave bound cluster remnants in Newtonian dynamics \citep{Brinkmann+2017}. We will summarise the results in Sec. \ref{summary}.

\section{Embedded clusters and gas expulsion}\label{ics}

\subsection{Models in Newtonian dynamics and in MOND}

MOND has been tested for decades on the scale of galaxies, and the flat rotation curves for galaxies at large radii can be reproduced and in fact predicted from the observed density profile of luminous matter \citep[e.g.,][]{Sanders_McGaugh2002,Famaey_McGaugh2012,Famaey_Binney2005,Famaey+2007} by the empirical proposal of \citet{Milgrom1983a} and \citet{BM1984} without dark matter particles. Moreover, on the scale of star clusters, a number of recent observations and simulations in the Milky Way suggest deviations from Newtonian dynamics. The line-of-sight (LoS) velocity dispersion profiles, $\lossig$, of these GCs are flat at large radii \citep{Scarpa+2007,Scarpa+2011,Scarpa_Falomo2010,Lane+2009,Lane+2010,Hernandez_Jimenez2012,Hernandez+2013,Durazo+2017}. In the case of MOND, the overall LoS velocity dispersion for an isolated and isotropic virialised system, $\sigma_{\rm LoS}^2=\frac{1}{3} \sigma^2=\frac{2}{9}(GM_{\rm b}a_0)^{1/2}$, where the baryonic mass is $M_{\rm b}$ \citep{Milgrom1994}. The $\lossig$ profile for a virialised and isotropic MOND system within an external field has been fitted by an analytic function in \citet{Haghi+2009,Kroupa+2018nature} and Haghi et al. (2019, submitted), where $R$ is the 2-dimensional radius on the projected plane.

On large scales, MOND is confronted with several problems: the possible requirement of additional dark matter on the scale of clusters of galaxies (e.g., the lensing data from clusters, \citealt{Natarajan_Zhao2008,Angus+2007}, the Bullet Cluster 1E0657-56, \citealt{Clowe+2006}, and the dark matter ring surrounding the rich galaxy cluster Cl0024+17, \citealt{Jee+2012}) and the incorrect amplitude for the mass function of galaxy clusters \citep{Angus_Diaferio2011,Angus+2013,Angus+2014}. In addition, There is an inconsistency between MOND predictions and the observations of the third peak of the acoustic power spectrum of the CMB \citep{Skordis+2006,Zhao2008,Angus2009}. 
MOND is a classical theory of gravitation and cannot readily be applied to cosmological problems without additional assumptions.

A remarkable success of MOND is that it precisely predicts the observed correlation between the baryonic matter and apparent dark matter in a galaxy. It has been pointed out that for example, the empirical mass-discrepancy-to-acceleration (MDA) correlation, cannot be exactly reproduced by the cold dark matter (CDM) hypothesis \citep{Sanders1990,McGaugh2004,Famaey_McGaugh2012,Wu_Kroupa2015}. Although there are several recent simulational results showing that a correlation between the baryonic and dark matter distribution in a galaxy can be found in the $\Lambda$CDM framework \citep{DiCintio+2016,Read+2016,Keller_Wadsley2017,Navarro+2017,Ludlow+2017}, the scatter of the relation from the $\Lambda$CDM simulations is generally much higher than that from the observations (see also \citealt{Wu_Kroupa2015}). Therefore, it is important to continue studying MOND as a competitive alternative to the significantly challenged \citep{Kroupa+2010,Kroupa2012,Kroupa2015,Famaey_McGaugh2012} hypothesis of CDM.
\subsubsection{$N$-body initial conditions and numerical tools}
The initial parameters for the models of the embedded star clusters are the same as those in \citet{Wu_Kroupa2017}. The models are spherically symmetric and isotropic, with a Plummer density distribution \citep{Plummer1911}:
\beq
\rho_{\rm b} (r)=\left(\frac{3\mb}{4\pi \rp^3} \right)\left(1+\frac{r^2}{\rp^2}\right)^{-5/2},
\eeq
where $\mb$ and $\rp$ are the total mass and the scale length of the Plummer model. Besides the same MONDian models presented in \citet[][models M1-M4 in Table \ref{plummer}]{Wu_Kroupa2017}, an equilibrium model (model N1) in Newtonian dynamics is constructed with the same density distribution as that of M1. We only construct one Newtonian model to compare with the MOND models, since the Poisson equation is linear and gravity is scale-free in Newtonian dynamics. The $N$-body initial conditions (ICs) are constructed using Lucy's method \citep{Lucy1974}. The parameters of the Newtonian and MONDian models are summarised in Table \ref{plummer}. There are $10^5$ equal-mass particles in each model. For the MOND models we used the same interpolating function, $\mu(X)= \frac{X}{1+X}$, as that in \citet{Wu_Kroupa2017}, where $X=\frac{|\vg|}{a_0}$ and $\vg$ is the gravitational acceleration in MOND.

Since the number of particles is large, two-body relaxation can be ignored in the simulations. The evolution of models after the gas expulsion is calculated by using the collision-less particle-mesh $N$-body code \emph{NMODY} \citep{nmody}. \emph{NMODY} calculates  gravitational accelerations and potentials by solving Poisson's equation in both standard Newtonian and Milgromian dynamics \citep{Nipoti+2007,Nipoti+2008,Nipoti+2011,Wu_Kroupa2013,Wu+2017}. A grid resolution of $n_r\times n_\theta \times n_\phi = 256\times 32 \times 64$ is chosen for the simulations, where $n_r,~n_\theta$ and $n_\phi$, respectively, are the number of cells in radial, polar and azimuthal dimensions. Here $n_r$ cells are used within a radial range of $0.15\pc$-$65.32\kpc$. The segmentation of grids is the same as that in Sec. 2.2 in \citet{Wu_Kroupa2017}. The global time steps are $dt=\frac{0.1}{\sqrt{\rm max |\dirv {\bf g}|}}$, which ensures that there are 10 time steps for a circular orbit in the densest region of a system. The models are evolved for about $2\Gyr$ after gas expulsion.

We emphasise that a direct $N$-body algorithm does not exist for MOND, which is a mean-field description, because the generalised Poisson equation in MOND is non-linear. Thus only two-body relaxation free, i.e., collision-less, systems can at present be simulated in MOND. Applying star clusters to test MOND \citep{Baumgardt+2005} may thus not be straightforward. The linear standard Poisson equation of Newtonian gravitation, on the other hand, has an excellent direct $N$-body algorithm \citep{Aarseth2003}.

\subsection{Gas expulsion in Newtonian dynamics and in MOND}\label{gas}
Extensive numerical simulations have been performed showing that there is a critical value of the SFE, $\epsilon_{\rm crit}\approx 0.3$, for a single and equilibrium embedded cluster to leave a bound remnant after gas expulsion in Newtonian dynamics \citep{Lada+1984,Goodwin1997,Kroupa+2001,Geyer_Burkert2001,Boily_Kroupa2003,Fellhauer_Kroupa2005,Baumgardt_Kroupa2007}. In MOND, $\epsilon_{\rm crit}$ is much lower, and can be down to $2.5\%$ for an initially diffuse embedded cluster \citep{Wu_Kroupa2017}. In order to compare the kinematics of the final star cluster, i.e., the self-bound virialised object (see Sec. \ref{kin}), in Newtonian dynamics and in MOND, we use $\epsilon=0.5$ and $0.4$ for the simulations in both dynamics. The above values of $\epsilon$ for an embedded cluster can leave a bound object after gas expulsion in both dynamics. Instantaneous gas expulsion is introduced by multiplying the mass of each particle by $\epsilon$ at the beginning of the simulations. 

Since NGC 2419 is $89~\kpc$ away from the Galactic centre, the current tidal force from the Milky Way disc can be ignored in Newtonian dynamics. In MOND, the strong equivalence principle is violated due to the existence of the external field from the Galaxy. It is therefore necessary to consider both the internal and external gravitational accelerations when studying the dynamics of a system.
At a Galactocentric distance of $89~\kpc$, the strength of the external field is $\approx 0.1a_0$ \citep{Wu_Kroupa2015}. Since the external field is much weaker than the internal field of the majority of NGC 2419\footnote{The internal gravitational acceleration of a system with a mass of $10^6\msun$ has been presented in Fig. 1 of \citet{Wu_Kroupa2013}. The internal acceleration approaches $0.1a_0$ at distances from the cluster centre larger than about $300~\pc$, which is more than $10\rh$ for NGC 2419.}, the MOND models are simulated in isolation. The external field effect can be included by a truncation radius, i.e., the virial radius $r_{\rm vir}$, 
 \beq\label{rvir} r_{\rm vir} \equiv \sqrt{GM_{\rm b,f}a_0}/g_{\rm ext},\eeq
for the phantom dark matter halo \citep[][]{Wu_Kroupa2015,Wu_Kroupa2017}. Here $g_{\rm ext}$ is the strength of the external gravitational acceleration. At the radius of $r_{\rm vir}$, the strength of the internal gravitational acceleration equals that of the external gravitational acceleration. Due to the external gravitational field, stars with sufficiently high energy are able to escape. The phantom dark matter halo is truncated at $r_{\rm vir}$, beyond which the external gravitational field from the Milky Way Galaxy dominates. $r_{\rm vir}$ is much smaller than the tidal radius in a MOND system \citep{Wu_Kroupa2017}. This indicates that the tidal field can be ignored in MOND. By the definition of $r_{\rm vir}$, the bound fraction of stars in a MOND final star cluster can be calculated from our simulations. The bound fraction of stars is the mass fraction of stars within $r_{\rm vir}$ in MOND, and within the tidal radius of the final star clusters in Newtonian dynamics.

\subsubsection{Radial-orbit instability and shape of the final GCs}
\begin{figure*}
\begin{centering}
\resizebox{14cm}{!}{\includegraphics{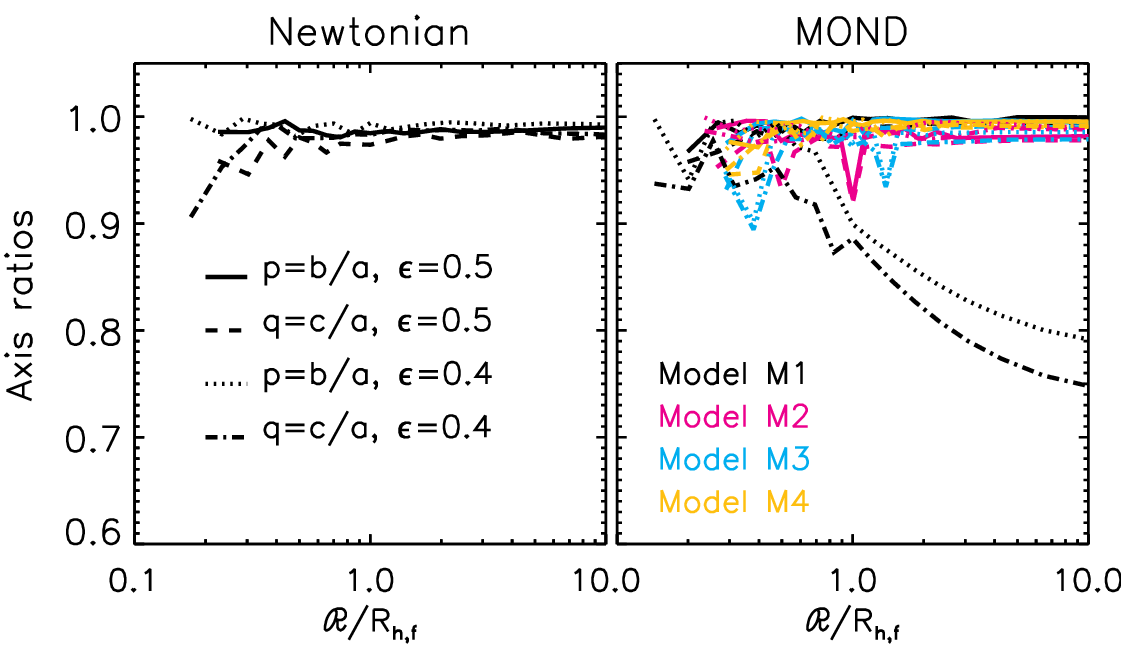}}
\caption{The axial ratios of the final star clusters undergoing gas expulsion in Newtonian dynamics (left panel) and in MOND (right panel). The four MOND models are differentiated by colours.}\label{ar}
\end{centering}
\end{figure*}

Since radial orbits are generated due to the shallower gravitational potential caused by gas expulsion, the star clusters might suffer from radial-orbit instability. For an initially spherical system, the radial-orbit instability alters the shape of the system into a triaxial morphology. Moreover, the velocity dispersion profile becomes more isotropic in the central region and more radially anisotropic in the outer region of the system \citep{Barnes+2005,Bellovary+2008}. The LoS velocity dispersion could differ along different projected angles of a star cluster, if the shape deviates significantly from a sphere. It is important to examine the radial-orbit instability of the star clusters undergoing gas expulsion. 

The radial-orbit instability can be characterised by an anisotropy parameter, $\zeta\equiv \frac{2K_{\rm r}}{K_{\rm \theta}+K_{\rm \phi}}$, where $K_{\rm r}, ~K_{\rm \theta}$ and $K_{\rm \phi}$ are the kinematic energy on the radial, arzimuthal and polar directions in a spherical coordinate system \citep{Polyachenko_Shukhman1981,Saha1992,Bertin+1994,Trenti_Bertin2006}. The radial-orbit instability appears in a spherical system when the value of $\zeta$ is larger than a critical value, $\zeta_{\rm crit}$. The values of $\zeta_{\rm crit}$ are in a range of around [1.6, 2.3] for spherical models with different distribution functions \citep[e.g.][]{May_Binney1986,Saha1991,Saha1992,Bertin_Stiavelli1989,Bertin+1994}. A spherical system is generically unstable when $\zeta \gtrsim 2.3$ in Newtonian dynamics. The radial-orbit instability of spherical Osipkov-Merritt MOND and Newtonian systems have been studied by \citet{Nipoti+2011}. The values of $\zeta_{\rm crit} \in [2.3,~2.6]$ in MOND systems and $\zeta_{\rm crit} \in [1.6,~1.8]$ in pure baryonic systems in Newtonian dynamics ($\xi_s$ in their table 1 is our $\zeta_{\rm crit}$). \citet{Nipoti+2011} demonstrate that a pressure-supported MOND model is more radially unstable than an equivalent Newtonian system where a dark matter halo reproduces the potential which is equivalent to a MOND potential. But a MOND model is more stable than a pure Newtonian model without any dark matter. This is evident from the values of the critical instability parameter. We have examined the $\zeta$ values of all the final star clusters, and found that $\zeta\in[1.10,~1.72]$ for all the simulations in Table \ref{plummer} (initial models $N1,~M1-M4$). This is consistent with the results of \citet{Nipoti+2011}, because none of the final star clusters have $\zeta$ exceeding the above range of values of $\zeta_{\rm crit}$. However, it is possible that the radial-orbit instability has been at work during the evolution of some of the simulated clusters. We address this question by calculating the shapes of all the final star clusters.

The shape of a final star cluster is described by the ratios of the principal axes, i.e., the intermediate-long axial ratio, $p$, and the short-long axial ratio, $q$. The $p$ and $q$ at a radius of $\rt=\sqrt{x^2+(y/p)^2+(z/q)^2}$ can be calculated from the diagonalised normalised moments of the inertia tensor of the particle system, with an initial guess of $p=q=1.0$. The procedure follows \citep[also see][]{Wu+2017}:
\begin{itemize}
\item 1. The ($I_{\rm xx}(\rt),~I_{\rm yy}(\rt),~I_{\rm zz}(\rt),~I_{\rm xy}(\rt),~I_{\rm xz}(\rt),~I_{\rm yz}(\rt)$) of the normalised moments of inertia tensor, $I(\rt)$, are computed for a particle system within $\rt$. Here \beq I_{\rm xx}(\rt)= \frac{\sum_i m_i(y_i^2+z_i^2)/\rt_i^2}{\sum_i m_i}.\eeq 
$I_{\rm yy}(\rt)$ and $I_{\rm zz}(\rt)$ are defined in the same way. The component 
\beq I_{\rm xy}(\rt)= \frac{\sum_i -m_ix_iy_i/\rt_i^2}{\sum_i m_i}.\eeq 
$I_{\rm xz}(\rt)$ and $I_{\rm yz}(\rt)$ are defined in the same way. $(x_i,~y_i,~z_i)$ are the Cartesian coordinate of the $i_{th}$ particle within $\rt$. The inertia tensor of each particle is normalised by its radius $\rt_i$. 
\item 2. The eigenvalues and eigenvectors of $I(\rt)$ are calculated by diagonalising the matrix of normalised inertia tensor. The eigenvalues are the new normalised $I_{\rm x'x'}(\rt),~I_{\rm y'y'}(\rt)$ and $I_{\rm z'z'}(\rt)$, where $(x',~y',~z')$ is the new reference frame which aligns with the principal axes of the particle system within $\rt$. The normalised scale lengths of the principal axes of the particle system, $a',~b',~c'$, can be obtained by
\bey
a'(\rt)=\sqrt{(I_{\rm y'y'}+I_{\rm z'z'}-I_{\rm x'x'})/2},\nonumber \\
b'(\rt)=\sqrt{(I_{\rm x'x'}+I_{\rm z'z'}-I_{\rm y'y'})/2},\\
c'(\rt)=\sqrt{(I_{\rm x'x'}+I_{\rm y'y'}-I_{\rm z'z'})/2}.\nonumber
\eey
The axial ratios of the particle system within $\rt$ are $p=b'/a'$ and $q=c'/a'$.
\item 3. The whole final star cluster is rotated to align with the principal axes using the rotation matrix which diagonalises the normalised moments of inertia. In the new reference frame, the radius $\rt$ is updated by using the new values of $p$ and $q$. 
\item 4. We repeat the procedure from step 1 to step 3 until the axial ratios converge in the $j_{th}$ loop, in which $\Delta p_j=(p_j- p_{j-1})/p_{j-1}<10^{-4}$ and $\Delta q_j=(q_j- q_{j-1})/q_{j-1}<10^{-4}$.
\end{itemize}

The axial ratios of the final star clusters as functions of rescaled radial distance, $\rt/\rhf$, are shown in Fig. \ref{ar}. In Newtonian dynamics (left panel), for the final star cluster with $\epsilon=0.5$, $p$ is very close to 1.0 with $2\%$ deviation, while $q$ is about 0.9 in the innermost region ($\rt<0.2\rhf$) and it grows quickly with the increase of radius. At a radius of $\approx 0.4\rhf$, $q\in [0.98, 1.0]$. Such a final star cluster is thus almost perfectly spherical. For the final star cluster with $\epsilon =0.4$, the axial ratios are very similar to those of the star cluster with $\epsilon=0.5$. Therefore, the Newtonian models are almost spherical at $r>0.4\rhf$. 

In MOND, the axial ratios of the final star clusters of models $M2-M4$ with $\epsilon$ of 0.4 and 0.5 behave very similarly to those in Newtonian dynamics. So do the axial ratios of the final star cluster of $M1$ with $\epsilon=0.5$. These MOND star clusters are almost spherical, especially in the outer regions with $\rt>\rhf$. The shape of the final star cluster of $M1$ with $\epsilon=0.4$ behaves very differently to the Newtonian models and the other MOND models. In this star cluster, $p$ drops from near 1.0 to 0.79 within $\rt\in [0.7,10.0]\rhf$, and $q$ declines from about 0.94 to 0.75 within $\rt \in [0.5,10.0]\rhf$. This is a triaxial star cluster, which implies that the radial-orbit instability occurs in the $M1$ model with $\epsilon=0.4$ after gas expulsion. $M1$ is the most massive model, and the dynamics is expected to be the most similar to that of the Newtonian model. But this is not the case for the process of gas expulsion. Initially, models $N1$ and $M1$ are very similar, since the MOND effect is weakest in $M1$. When the gas is expelled from the star clusters with $\epsilon=0.4$, a large number of stars are moving along extremely radial orbits in both star clusters. These stars escape from the star cluster in Newtonian dynamics, but most of them remain bound in MOND within the truncation radius, $r_{\rm vir}$, where the strength of the internal and external field are comparable (see the bound fraction in Table \ref{plummer}). These stars account for the radial-orbit instability of the final star cluster of $M1$ in MOND. For the star cluster evolved from $M1$ with $\epsilon=0.5$, the potential does not change as much as that with $\epsilon=0.4$. For the $M2-M4$ models, since they are less massive than $M1$, the change of the MOND potential is smaller than that of Newtonian potential after gas expulsion. Fewer extremely radial orbits are generated in these MOND models. As a result, these final star clusters are nearly spherical. The radial-orbit instability in these models agree with the previous analysis of the instability parameter, $\zeta$.


We shall further study the surface density and kinematics within the final clusters in the following subsections. All the final star clusters are rotated to align with their principal axes. 

\subsubsection{Mass distribution after gas expulsion}
\begin{figure*}
\begin{centering}
\resizebox{14cm}{!}{\includegraphics{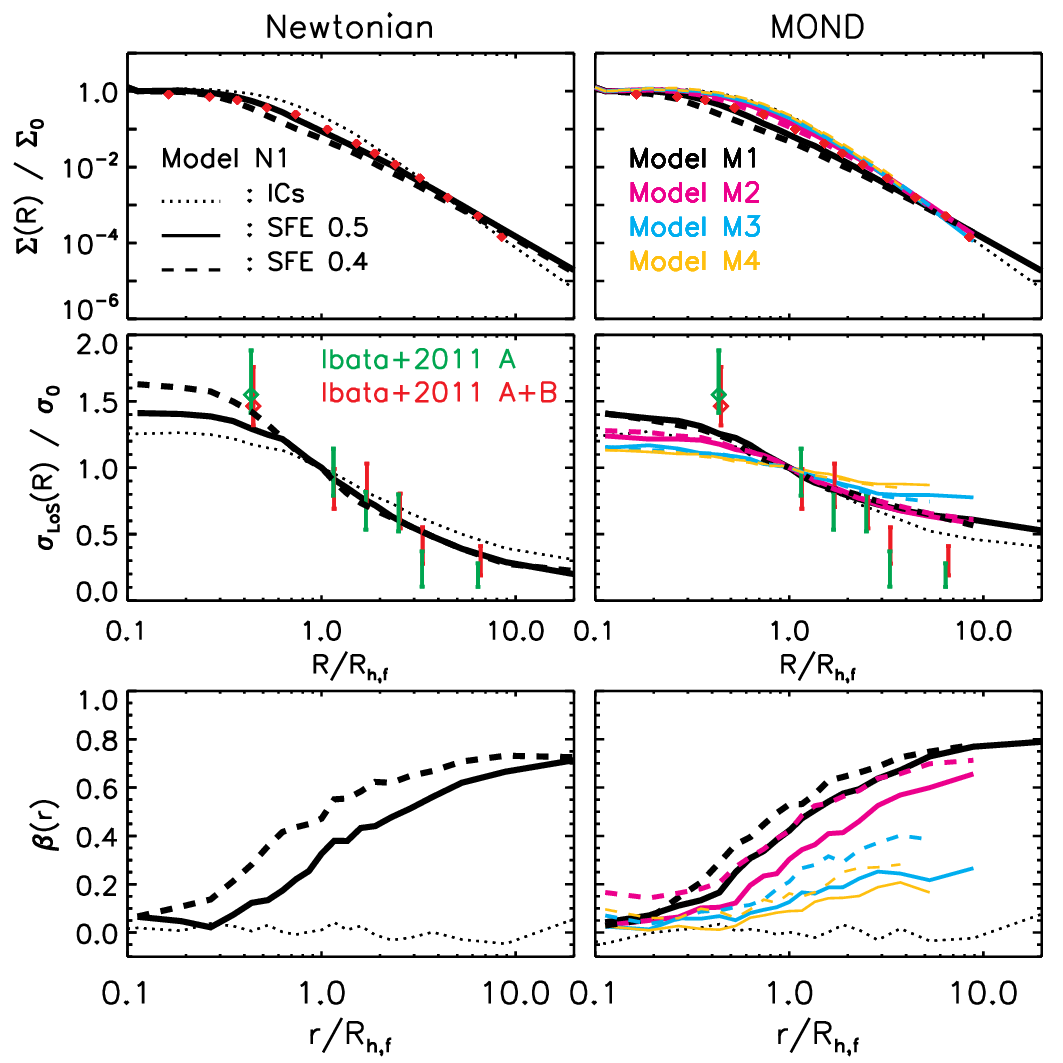}}
\caption{Upper panels: the surface density profiles, $\Sigma(R)$, at time $T=2\Gyr$ and initially, normalised by the effective surface density, $\Sigma_{\rm 0,f}\equiv \frac{M_{\rm b,f}}{2} \rhf^{-2}$ and $\Sigma_0\equiv \frac{M_{\rm b}}{2} \rh^{-2}$, respectively. The red symbols with error bars (The error ranges are quite small for the observed surface density, especially on the logarithmic scale. The error bars cannot be distinguished in the figure.) show the observed density profile of NGC 2419 with the same normalisation. Middle panels: the LoS (i.e., one-dimensional) velocity dispersion (Eq. \ref{sigmalos}) profiles, $\lossig$, of the surviving star clusters at $T=2\Gyr$ and initially, normalised by the LoS velocity dispersion at $\rhf$ and $\rh$, i.e., $\sigma_{\rm 0,f}$ and $\sigma_{\rm 0}$, respectively. The radii are in units of $\rhf$ (and $\rh$ for the ICs). The data with error bars are the normalised observed $ \sigma_{\rm LoS}$ in \citet{Ibata+2011a} for the cluster NGC 2419. Lower panels: the anisotropy profiles for the models. Different line types are used to differentiate the ICs for embedded clusters (dotted), the final star clusters undergoing gas expulsion with an original $\epsilon$ of $0.5$ (solid) and of $0.4$ (dashed). The left panels are for models in Newtonian dynamics and the right panels are for models in MOND.}\label{newtkin}
\end{centering}
\end{figure*}

At first, the major axis, i.e., $x$-axis, is chosen as the direction of the line-of-sight.
The surface density distributions, $\Sigma(R)$, are presented in the upper panels of Fig. \ref{newtkin} for the Newtonian (upper left panel) models and the MONDian (upper right panel) models, where $R$ is the circular radial distance on the projected plane and the binning details are presented in Sec. \ref{kin}. The $\Sigma(R)$ profiles are rescaled by the central surface densities, i.e., the value of $\Sigma$ at $R=0$. The values of the central surface densities are shown in Table \ref{plummer}, where $\Sigma_0$ is for the ICs and $\Sigma_{\rm 0,f}$ is for the final star clusters.
After gas expulsion, the rescaled surface density profiles are very similar to the original rescaled Plummer profile in both Newtonian gravity and in MOND, which agrees well with the results in \citet{Boily_Kroupa2003}.

The V-band surface brightness profile of NGC 2419 measured by \citet[][their Table 2]{Ibata+2011a} with errors is converted to $I_{\rm V}$ in units of $\lsun \pc^{-2}$. A constant mass-to-light ratio, $\Upsilon$, is assumed here, as has been used for the modeling of NGC 2419 in \citet{Ibata+2011a,McGaugh2012}, and the normalised surface density profile, 
\beq \Sigma/\Sigma_0(\rm NGC 2419) (R) =\frac{I_{\rm V}(R)\cdot \Upsilon}{I_{\rm V, \rh=19pc}\cdot \Upsilon},
\eeq 
is shown in the upper panels of Fig. \ref{newtkin} (red symbols). The observed surface density profile of NGC 2419 can be well reproduced by the rescaled Plummer's profile. In general, the surface density profiles in both Newtonian dynamics and in MOND appear to be similar to that of the observed surface brightness profiles.
The rescaling at the central regions of the star cluster makes it possible to directly compare the particle models and the observed surface brightness profile. One does not need to assume a mass-to-light ratio, which might introduce new uncertainties, for NGC 2419 when comparing the observations with models.

\begin{figure*}
\begin{centering}
\resizebox{14cm}{!}{\includegraphics{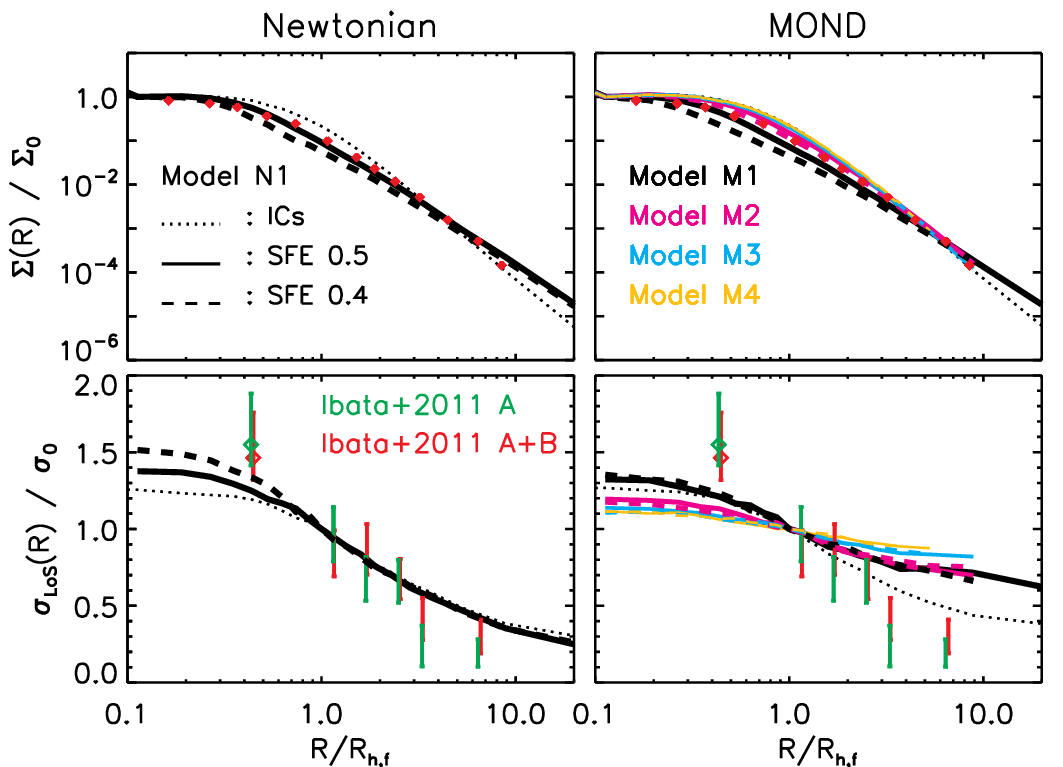}}
\caption{Upper panels: the surface density profiles of the final star clusters with $y$-axis as the LoS direction.
Lower panels: the normalised LoS velocity dispersion profiles along the $y$-axis, $\lossig$, of the surviving star clusters at $T=2\Gyr$ and initially. Lower panels: the anisotropy profiles for the models.}\label{newtkin_losy}
\end{centering}
\end{figure*}

\begin{figure*}
\begin{centering}
\resizebox{14cm}{!}{\includegraphics{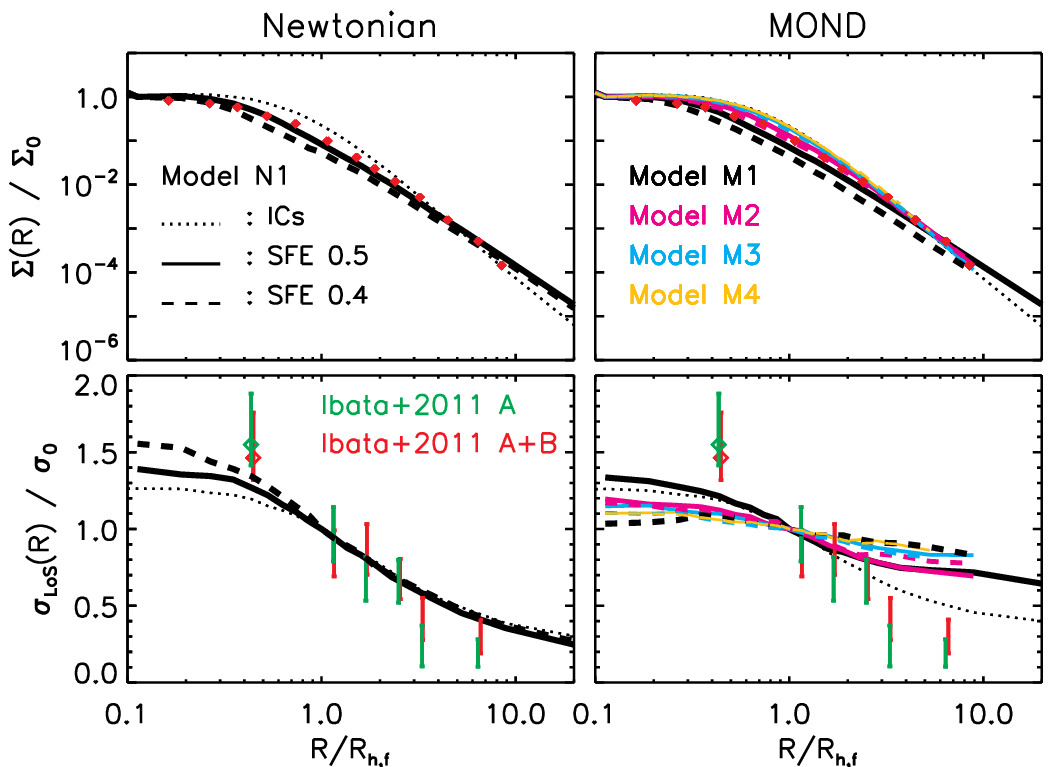}}
\caption{Upper panels: the surface density profiles of the final star clusters with $z$-axis as the LoS direction.
Lower panels: the normalised LoS velocity dispersion profiles along the $y$-axis, $\lossig$, of the surviving star clusters at $T=2\Gyr$ and initially. Lower panels: the anisotropy profiles for the models.}\label{newtkin_losz}
\end{centering}
\end{figure*}

Since the final star clusters are not spherical in the central regions in both Newtonian and Milgromian dynamics, we also present the surface density profiles by choosing $y$-axis and $z$-axis as the LoS directions. The normalised surface density profiles on the $x-z$ plane and $x-y$ plane are shown in the upper panels of Figs. \ref{newtkin_losy}-\ref{newtkin_losz}. We find that the normalised surface density profiles are very similar to the profiles shown in the upper panels of Fig. \ref{newtkin}.

\subsection{Kinematics}\label{kin}
The middle panels of Fig. \ref{newtkin} display the LoS ($x$-axis direction) velocity dispersion, i.e.,
\beq\label{sigmalos}
\sigma_{\rm LoS}^2(R)= {1\over N_{\rm bound}}\sum_{i=1}^{N_{\rm bound}}(v_{\rm x,i}-{\bar v_{\rm x}})^2,
\eeq
where $N_{\rm bound}$ is the number of bound particles in the star cluster after gas expulsion and ${\bar v_{\rm x}}$ is the mean of the velocity component $v_{\rm x}$. Technically, different binning methods of the radius lead to slightly different profiles of $\lossig$. The radius $R$ is binned following the equation:
\beq
R_j=\rhf \tan [(j+0.5)0.5\pi/(n_R+1)],
\eeq
where $n_R=20$ and $j=1,~2,~3,...,n_R$. The value of $\sigma_{\rm LoS}(R_{j+1})$ is the LoS velocity dispersion of particles within $R_{j}$ and $R_{j+1}$, and the related projected radius is the mean projected radius of particles within $R_j$ and $R_{j+1}$.

The $\lossig$ profiles are normalised by the effective velocity dispersion, i.e., velocity dispersion interpolated at the projected half-mass radius (denoted as $\rh$ for the ICs and $\rhf$ for the 2 Gyr old star clusters after undergoing sudden gas expulsion initially at time T=0), $\sigma_{\rm 0}\equiv \sigma_{\rm LoS}|_{\rm r=\rh}$ for the initial models and $\sigma_{\rm 0,f}\equiv \sigma_{\rm LoS}|_{\rm r=\rhf}$ for the final star clusters, respectively. The observed LoS velocity dispersion profile for NGC 2419 \citep{Ibata+2011a} is listed in Table \ref{kin2419} and plotted in the middle panels for comparison, where the normalisation parameter is $\sigma_{\rm Los}$ at the projected half-mass radius of NGC 2419 (i.e., 19pc). Moreover, the anisotropy profiles of the velocity dispersion,
\beq \beta(r)\equiv 1-(\sigma_\theta^2+\sigma_\phi^2)/2\sigma_r^2,\eeq
truncated at the virial radius, $r_{\rm vir}$ \citep{Wu_Kroupa2017}, are shown in the lower panels of Fig. \ref{newtkin}, where $\sigma_r$, $\sigma_\theta$, and $\sigma_\phi$ are the components of the velocity dispersions in spherical coordinates $(r,~\theta,~\phi)$. The $\beta(r)$ profiles are isotropic for all the ICs. 

In addition, the crossing time of the final star clusters is defined as 
\beq \tcr= \rhf / \sigma_{\rm 0,f}. \eeq
The simulation time of $2~\Gyr$ is always longer than $300~\tcr$ of the final star clusters in both dynamics listed in Table \ref{plummer}. The final star cluster models are fully re-virialised with a SFE between 0.4 and 0.5.

\subsubsection{GCs in Newtonian dynamics}

The Newtonian models predict larger values of $\sigma_{\rm LoS}/\sigma_{\rm 0,f}$ in the central regions for the final star clusters. The normalised $\lossig$ is larger for the N1 model with a smaller value of $\epsilon$. However, the normalised $\lossig$ profiles cannot be distinguished for N1 models with different $\epsilon$ in the outer region where $R>\rhf$. The $\lossig$ profiles for the Newtonian models agree well with those of the two observational samples (samples $A$ and $A+B$) of stars for NGC 2419 in \citet{Ibata+2011a}. The data are binned with an equal number of stars as the radius increases, and the radius of a bin is taken as the average radius of the observed stars in the binning range. 
Since the normalised $\lossig$ profiles for the final model star clusters are almost indistinguishable from the radius of the innermost bin of the observed data, any star cluster with $\epsilon \leq 0.5$ undergoing sudden gas expulsion should leave the same normalised $\lossig$ profile in Newtonian dynamics. For the case of a star cluster with $\epsilon < 0.4$, a more rapidly declining $\lossig$ profile with increasing $R$ for radii smaller than $\rhf$ will be left, due to the fact that more radial orbits are generated through a larger fraction of mass loss. Observations show that the SFE is less than $0.3$ for the dense clumps in giant molecular clouds \citep{Lada_Lada2003,Higuchi+2009,Kainulainen+2014,Megeath+2016}. Consequently, an embedded cluster model with $\epsilon>0.5$ disagrees with the observational constraints and we do not simulate models with $\epsilon>0.5$.

The $\beta(r)$ profiles (lower left panel) show that the final star clusters are radially anisotropic, especially in the outer regions of the star clusters where $R>\rhf$. The radial anisotropy increases when $\epsilon$ is smaller. The velocity dispersion for the final star cluster with $\epsilon=0.4$ is extremely radially anisotropic. For a model with $\epsilon=0.5$, $\beta(\rhf)\approx 0.4$, while for $\epsilon=0.4$, $\beta(\rhf)\approx 0.6$. These $\beta(r)$ profiles result in almost indistinguishable normalised $\lossig$ profiles, which agree well with the observed data of NGC 2419, especially for the star sample A+B in \citet{Ibata+2011a}. The process of gas expulsion predicts that the $\beta(r)$ profiles for sufficiently isolated GCs are strongly radially anisotropic. Note that the initial embedded star clusters are isotropic. The $\beta(r)$ profiles of the final star clusters  might be less radially anisotropic if the embedded star clusters are strongly tangentially anisotropic and then undergo gas expulsion. 

In addition, the normalised LoS velocity dispersion along the $y$-axis and $z$-axis for the same final star clusters are shown in the lower left panels of Figs. \ref{newtkin_losy}-\ref{newtkin_losz}. The LoS velocity dispersion profiles are almost the same along the three principal axes' directions. This further confirms again that gas expulsion can naturally explain the kinematics of NGC 2419, without introducing any assumptions on special directions as the line-of-sight.

Other physical processes, such as stellar evolutionary mass loss and two-body relaxation in a star cluster, will further drive the GCs to become more radially anisotropic. In the former case, the mass of a GC is further reduced and the gravitational potential becomes shallower. More stars move on elongated orbits in the evolving potential. In the latter case, the GC expands in the outer region and collapses in the core region. More radial orbits are expected in this process. However, a systematic study of the impact of these physical processes on the kinematics of GCs is still lacking. 
In addition, the tidal field from the Galaxy might allow the stars with high energy, i.e., stars on radial orbits, to escape. As a result, the velocity dispersion becomes less radially anisotropic. However, our models represent star clusters in the outer halo of a galaxy where the tidal field can be ignored. Therefore, a rapidly declining $\lossig$ profile is expected for GCs, especially for the GCs in the outer halo of a galaxy. Given that gas expulsion is physically realistic for all the GCs \citep[see][]{Marks+2012,Zonoozi+2011,Zonoozi+2014,Zonoozi+2017,Haghi+2015}, it is necessary to consider this process when one uses tidal heating models to explain the existence of observed flat $\lossig$ profiles at large radii for the Galactic GCs \citep{Scarpa+2007,Scarpa+2011,Scarpa_Falomo2010,Lane+2009,Lane+2010,Hernandez_Jimenez2012,Hernandez+2013,Durazo+2017}. However, the process of gas expulsion has been ignored in the above existing studies, therefore an initially strongly radially anisotropic velocity dispersion has not yet been taken into account. The details on the $\lossig$ and $\beta(r)$ profiles changed by stellar evolutionary mass loss, two-body relaxation and the tidal field in the inner region of the Galaxy under the consideration of gas expulsion in Newtonian dynamics, will be systematically studied in future projects by applying direct $N$-body simulations. 

\subsubsection{Newtonian models for NGC 2419}
The surface density and velocity profiles shown in Fig. \ref{newtkin} are normalised by constants. 
Therefore, it is possible to reproduce the surface density profile and the LoS velocity profile of NGC 2419 in physical units by choosing specific normalisation parameters for the initial models. 

The distance modulus of NGC 2419, $(m-M)_{\rm 0}=19.71\pm 0.08$ \citep{DiCriscienzo+2011}. The extinction is $E(B-V)=0.08\pm0.01{\rm mag}$ \citep{DiCriscienzo+2011}. The Solar absolute V-band magnitude is $M_{\rm V,\odot}=4.83$. The observationally determined V-band mass-to-light ratio of NGC 2419, $\Upsilon=M/L_{\rm V}$, lies in a wide range of $[1.2,~2.55]$ \citep{Bellazzini+2012,Ibata+2011a,Baumgardt+2009}. The $\Upsilon$ obtained from dynamical models are $1.9\pm 0.16$ \citep{Ibata+2011a} and $2.05\pm 0.5$ \citep{Baumgardt+2009}, respectively. On the other hand, the $\Upsilon$ derived from stellar evolution models is $1.2-1.7$ and the best fitting value is $1.5\pm 0.1$ with $1\sigma$ error \citep{Bellazzini+2012}.

\begin{figure}
\resizebox{8.5cm}{!}{\includegraphics{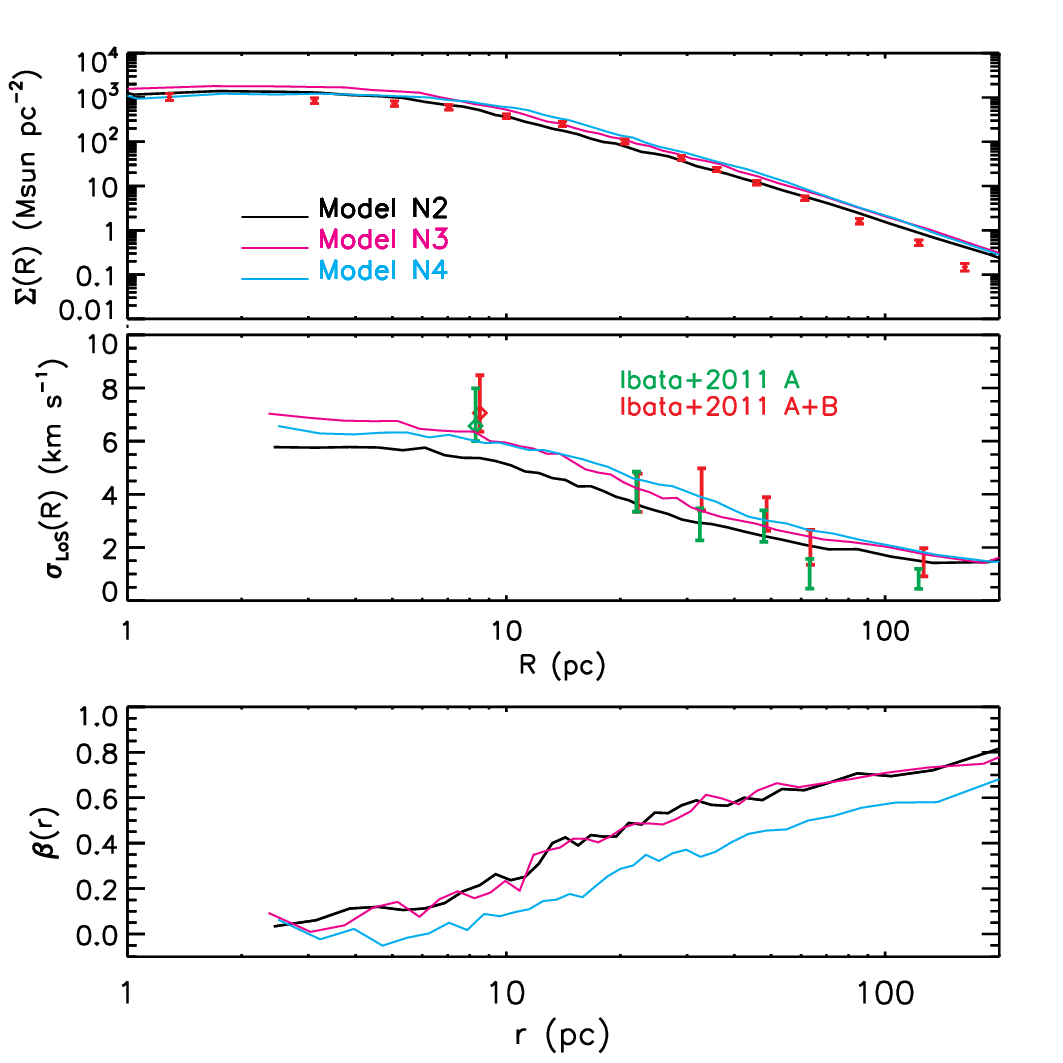}}
\caption{The Newtonian models (Table \ref{plum2419}) to reproduce the projected density (upper panel), LoS velocity dispersion profile $\lossig$ (middle panel) and anisotropy profile $\beta(r)$ (lower panel) of NGC 2419. }
\label{newtkin2419}
\end{figure}

To reproduce the mass, size and stellar velocities of NGC 2419, the initial mass and Plummer radius of the embedded cluster models need to be rescaled \citep[the method is described by ][]{Madejsky_Bien1993}. We present three sets of embedded cluster models which are possible to reproduce the mass distribution and kinematics of NGC 2419 in Table \ref{plum2419} and in Fig. \ref{newtkin2419}. We find that the surface density and LoS velocity dispersion profiles of the final star clusters evolving from models $N3$ and $N4$ agree the best with those of NGC 2419. Model $N2$ predicts final star clusters with a lower central velocity dispersion profile, which does not agree with the observed $\lossig$ data within the $1\sigma$ error range. Moreover, the overall masses of the three models are within $7.48\times 10^5$-$1.08\times 10^6\msun$, and they agree with the dynamical mass suggested by \citet{Ibata+2011a} and \citet{Baumgardt+2009}. The mass-to-light ratios, $\Upsilon$, of the models are within $1.70$-$2.46$, which are consistent with the observational V-band $M/L_{\rm V}$  \citep{Bellazzini+2012,Ibata+2011a,Baumgardt+2009}. Consequently, it is possible to reproduce the mass distribution and kinematics of NGC 2419 by the mechanism of gas expulsion in Newtonian dynamics.

\begin{table}
\begin{center}
  \caption{Parameters of ICs for the embedded (pre-gas expulsion) clusters with a Plummer density profile to reproduce the surface density and LoS velocity profiles of NGC 2419 in Newtonian dynamics: the $1_{\rm st}$ through $4_{\rm th}$ columns are the ID of the ICs, the total mass, $M_{\rm b}$, the Plummer radius, $\rp$ and the half-mass radii, $\rh$, of the ICs. The $5_{\rm th}$ through $7_{\rm th}$ columns are the star formation efficiency, $\epsilon$, the half-mass radii of the final star clusters after undergoing gas expulsion ($\rhf$) and the total mass of the final star clusters within the tidal radius, $M_{\rm f}$. The mass-to-light ratio (V-band) of the models are listed in the $8_{\rm th}$ column.}\vskip 0.30cm
\begin{tabular}{lccccccccccccc}
\hline
ICs& $M_{\rm b}$ & $\rp$ & $\rh$ & $\epsilon$ &$\rhf$ &$M_{\rm f}$& $\Upsilon$  \\
& $(\msun)$ & $(\pc)$& $(\pc)$ &  &$(\pc)$ & $(10^5\msun)$ & \\
\hline
N2& $5\times10^6$ &3.25& 3.25 & 0.4 & 18.0 & 7.48 & 1.70\\
N3 & $7\times10^6$ &3.25& 3.25 & 0.4 & 17.5 & 10.29 & 2.34 \\
N4 & $3\times10^6$ &6.50& 6.50 &0.5& 18.5 & 10.83 & 2.46\\
\hline
\end{tabular}
\label{plum2419}
\end{center}
\end{table}

\subsubsection{GCs in MOND}
In the case of MOND, the normalised $\lossig$ profiles along the $x$-axis (middle right panel of Fig. \ref{newtkin}) for the final star clusters decrease more slowly compared to the Newtonian models. The normalised $\lossig$ profile of the most compact model, i.e., the final star cluster evolved from the embedded cluster model M1, declines most steeply with radius. For this embedded cluster, when $\epsilon$ is larger (i.e., $\epsilon=0.5$), $\sigma_{\rm LoS}/\sigma_{\rm 0,f}$ is smaller in the centre and is slightly larger at large radii compared to that in Newtonian dynamics. The $\beta(r)$ profiles for the final star cluster of $M1$ with $\epsilon=0.4$ and $0.5$ are the most radially anisotropic. We find that the $\beta(r)$ profile for the final star cluster of model M1 is more radial than that of model $N1$ for the same SFE. Is this in contrast to the fact that the $\lossig/\sigma_{\rm 0,f}$ profile of the final star cluster of M1 does not decline with radius as fast as that of model $N1$? The $N1$ and $M1$ models are the most massive models in Newtonian dynamics and in MOND. The $\lossig/\sigma_{\rm 0,f}$ and $\beta(r)$ profile of the final M1 star cluster is therefore expected to be the most similar to that of the N1 model. As aforementioned, the more radially anisotropic velocity dispersion in MOND for the final M1 star cluster comes from a larger fraction of bound stars (see Table \ref{plummer}), including some extremely radial orbits which are escaping stars in N1 but are bound in M1. The very radial orbits only exist in the very compact and massive MOND models such as in M1. In this model, the potential is quasi-Newtonian at the beginning and evolves to an intermediate-MOND state after gas expulsion. More extremely radial orbits can be generated and also bound to the system. However, since the gravitational potential is deeper for the final star cluster of M1, all of the components of velocity dispersion ($\sigma_{\rm r},~\sigma_{\rm \theta}$ and $\sigma_{\rm \phi}$) are larger than those in Newtonian dynamics. This explains why the stronger radial anisotropy for the final star cluster of $M1$ cannot reproduce a rapidly declining $\lossig/\sigma_{\rm 0,f}$ profile like that of model $N1$.

When the embedded clusters become less compact, the slopes of the normalised $\lossig$ profiles decline with $R$ more gently. The values of $\lossig/\sigma_{\rm 0,f}$ for all the final star clusters are smaller in the centres and larger at large radii, compared to NGC 2419. The innermost bin of the observed normalised $\lossig$ is above the profile predicted by models and the last two bins of the observed data are below all the predicted $\lossig/\sigma_{\rm 0,f}$ curves in MOND. If MOND as a mean-field theory is applicable in star clusters, this indicates that gas expulsion in MOND is not the sole mechanism for the observed kinematics of NGC 2419. That is why the $\lossig/\sigma_{\rm 0,f}$ profiles of the models undergoing gas expulsion appear very different from those in \citet{Ibata+2011a}, where the models are constructed initially in equilibrium in the framework of MOND.

The LoS velocity dispersion along other principal axes, i.e., $y$- and $z$-axis, are illustrated in the lower right panels of Fig. \ref{newtkin_losy} and Fig. \ref{newtkin_losz}, respectively. For the final star clusters evolved from $M1$, the central $\sigma_{\rm LoS}/\sigma_{\rm 0}$ profiles along the $y$- and $z$-axis are smaller than those along the $x$-axis for both values of $\epsilon$, especially for the $\epsilon=0.4$ model along the $z$-axis. Furthermore, in the outer region where $R=10~\rhf$, the values of $\sigma_{\rm LoS}/\sigma_{\rm 0}$ are larger along the $y$- and $z$-axis than those along the $x$-axis. In short, the normalised LoS velocity dispersion profiles are flatter along the intermediate and the short axes. This is related to the triaxial shape of the final star cluster. For other models, the $\sigma_{\rm LoS}/\sigma_{\rm 0}$ profiles along the $y$- and $z$-axis are very similar to those along the $x$-axis.

Owing to the deeper gravitational potential in MOND, for a given $\epsilon$, a final star cluster does not expand as much as in Newtonian dynamics. The stellar orbits in the final star clusters are less elongated. Therefore, the anisotropy profiles are less radial in MOND compared to Newtonian dynamics. Moreover, thanks to the fact that the effective mass of phantom dark matter is larger, relative to the baryonic mass, for the more diffuse systems, the size expansion is smaller for these models (see \citealt{Wu_Kroupa2015} and \citealt{Wu_Kroupa2017}). Fewer radial orbits are generated in the more diffuse systems. This results in a more slowly declining $\lossig$ profile and a less radially anisotropic $\beta(r)$ profile for the more diffuse systems. The $\beta(r)$ profiles in the lower right panels of Fig. \ref{newtkin} confirm this. 

On the other hand, the proper motion for the Virgo stellar stream (VSS) indicates that NGC 2419 may be associated with the orbit of the VSS, which is a highly eccentric Galactic orbit with a pericentre of $\approx 11~\kpc$ \citep{Casetti-Dinescu+2009}. The external field effect from the Milky Way might account for the formation of an extreme radial anisotropy for NGC 2419 in MOND. That is, when a star cluster is moving on an eccentric Galactic orbit inwards and outwards, the strength of the external field from the Galaxy increases and decreases, respectively. As a result, the star cluster expands and contracts on such an orbit. Radial orbits in the star cluster are generated in this process, which might give rise to a radial anisotropy. The external field for outgoing orbits has been studied in \citet{Wu_Kroupa2013}. More sophisticated models including the effects of space-varying external field and gas expulsion will be studied for NGC 2419 in the future. 

Moreover, since the $\lossig$ profiles for the final star clusters are not unique in MOND, it is possible to reproduce the observed flat $\lossig$ profiles at large radii for distant GCs by using initially diffuse embedded cluster models and small values of $\epsilon$ undergoing gas expulsion. Such diffuse initial conditions would result from star formation throughout a giant molecular cloud, perhaps comparable to the Cyg OB2 association \citep{Knodlseder2000}. Distant GCs with moderate or mild radial anisotropy are thus allowed to exist in MOND. In contrast, all the GCs are expected to be strongly radially anisotropic after gas expulsion in Newtonian dynamics, unless another physical mechanism is taken into account.

To summarise, gas expulsion alone does not account for the observed velocity dispersion profile for NGC 2419 in MOND. Other physical mechanisms, such as, the time-varying external field effect, reproduce the $\lossig$ profile for NGC 2419 \citep{Wu_Kroupa2013}. However, star clusters surviving gas expulsion in MOND can have a slowly declining $\lossig$ profile with a mild or moderate radial anisotropy. These models can explain the observed flat $\lossig$ profiles at large radii for the distant Galactic GCs, while flat $\lossig$ profiles cannot be reproduced by gas expulsion-only models in Newtonian dynamics. One has to invoke other physical mechanisms to account for these profiles within the framework of Newtonian gravity.

\section{MOND models with ultra-low SFE}\label{ultralow}
\begin{figure}
\resizebox{8.5cm}{!}{\includegraphics{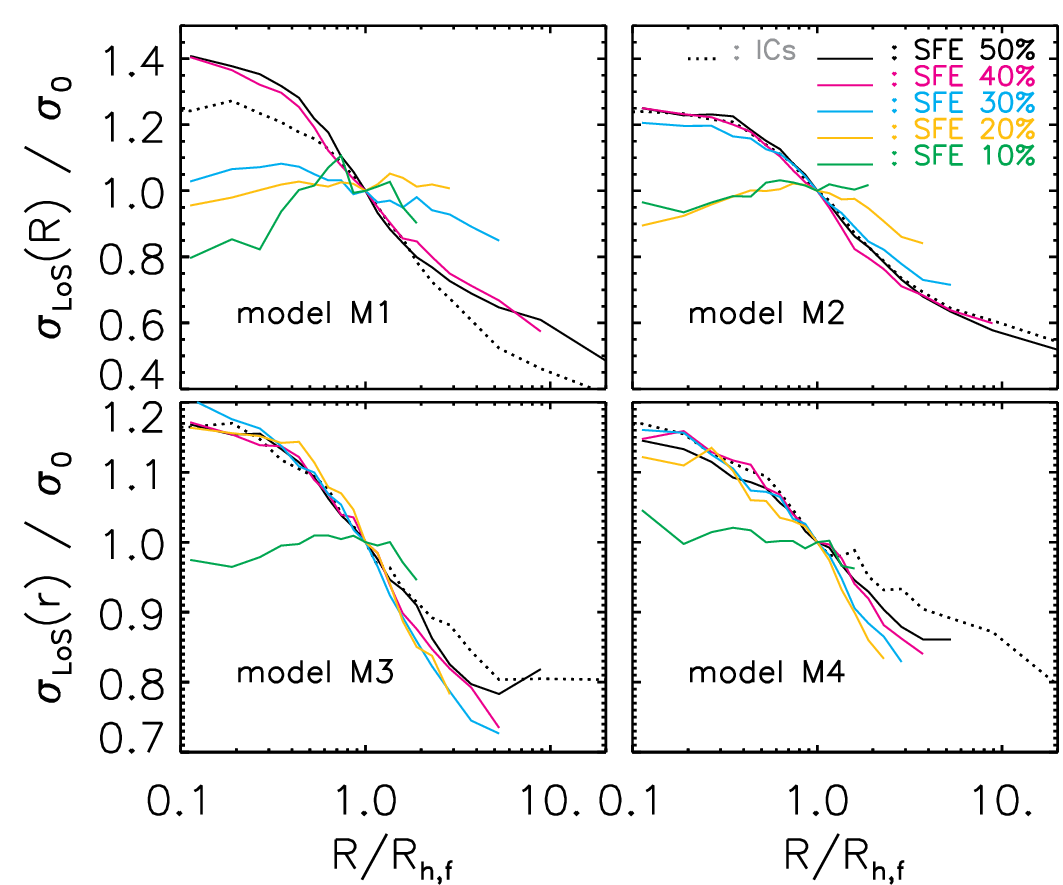}}
\caption{The LoS velocity dispersion profiles $\lossig$ of the ICs and the surviving star clusters for $\epsilon \leq 0.5$ evolved for $2\Gyr$ after gas expulsion, normalised by the LoS velocity dispersion at $\rh$ and $\rhf$, i.e., $\sigma_0 \equiv \sigma_{\rm LoS}|_{\rm r=\rh}$ and $\sigma_{\rm 0,f} \equiv \sigma_{\rm LoS}|_{\rm r=\rhf}$, respectively. The radii are in units of $\rhf$ (and $\rh$ for the ICs).}
\label{vlos}
\end{figure}

In Newtonian dynamics, the anisotropy profile of a final star cluster is more radially anisotropic when the SFE is lower for the same embedded cluster model. The question whether a smaller $\epsilon$ would make the $\lossig$ profile of a MOND model more compatible with the data from NGC 2419 then arises. Since models with ultra-low SFE can leave bound remnants after sudden gas expulsion in MOND \citep{Wu_Kroupa2017}, we shall further study the $\lossig$ and $\beta(r)$ profiles for such systems. Such low SFEs are typical for whole molecular clouds which spawn an OB association sufficiently massive to remain bound in MOND. The values of $\epsilon$ are chosen to be $30\%,~20\%$ and $10\%$ in the MOND models. In Newtonian dynamics, an embedded cluster with such values of SFE cannot survive sudden gas expulsion. For comparison, the embedded cluster models with $\epsilon$ of $0.5$ and $0.4$ are plotted in Fig. \ref{vlos} as well. We now make a pure model analyses by tuning $\epsilon$.

The normalised $\lossig$ profiles are shown in Fig. \ref{vlos}. In model M$1$, $\lossig$ of the star clusters surviving with $\epsilon$ of $50\%$ and $40\%$ have larger normalised values in the centres, i.e., $\sigma_{\rm LoS}/\sigma_{\rm 0,f}\approx 1.4$ at $0.1~\rhf$, compared to the initial models, i.e., $\sigma_{\rm LoS}/\sigma_0 \approx 1.2$. The velocity dispersion profiles of model M$1$ with these SFEs decline slowly within $0.5~\rhf$ and then drop rapidly with radius within $2~\rhf$. The normalised $\lossig$ profiles for these models decrease more slowly than those of the ICs again at radii $R>2\rhf$. Consequently, a compact embedded cluster with a large value of $\epsilon$ may result in a bound star cluster with a rapidly declining $\lossig$ profile. The velocity dispersion profiles are almost flat at all radii for $\epsilon$ of $30\%$ and $20\%$, and there is an increasing normalised $\lossig$ profile with increasing $R$ for model M$1$ when $\epsilon$ decreases down to $10\%$. The reason is that the bound mass for this surviving star cluster is small ($\approx 6\times 10^4\msun$, see \citealt{Wu_Kroupa2017}), while the size of the star cluster is as large as $\rhf\approx 200\pc$. The final ``star cluster'', formed from a giant molecular cloud complex with an overall SFE typical of molecular clouds ($\epsilon \in [0.2\%,~20\%]$ and with a luminous-weighted average of $8\%$ for a large sample of GMCs in the Galaxy, see \citealt{Murray2011}), is an ultra-diffuse system, and accordingly, the relaxation time scale is longer than $2~\Gyr$. The growing $\lossig$ profile with $R$ indicates a not yet fully virialised surviving star cluster. This agrees with the evolution of Lagrangian radii for this star cluster \citep{Wu_Kroupa2017}.

The values of the central normalised velocity dispersion, $\sigma_{\rm LoS}(R \approx 0.1\rhf)/\sigma_{\rm 0,f}$, are smaller when the models are less massive or more diffuse. 
Noteworthily, the normalised $\lossig$ profiles of models M$2$-M$4$ almost overlap those of the ICs. This indicates that although the radii of the clusters and the values of $\sigma_{\rm 0,f}$ evolve after gas expulsion, the radial distribution of velocity dispersion of these models does not change. The shapes of $\lossig$ of models M$2$-M$4$ with $\epsilon=0.1$ are flat, which is similar to model M$1$. In addition, when the model is more compact, the overall $\lossig$ profile is flat for a larger value of $\epsilon$. For instance, model M1 with $\epsilon \leq 0.3$ leaves a flat $\lossig$ profile, while in models M$3$ and M$4$, the $\lossig$ profiles are flat only when $\epsilon$ is reduced down to $0.1$. This is also because, i) the size expansion is the largest for model M$1$ and the smallest for models M$3$-M$4$ with the same value of $\epsilon$ and ii) the bound fraction is the smallest for model M$1$ and the largest for models M$3$-M$4$ for the same SFE \citep{Wu_Kroupa2017}. The virialisation timescales for star cluster models with an ultra-low SFE are longer than the simulation time of 2 Gyr. Hence the flat $\sigma_{\rm LoS}$ profiles might be related to the not yet virialised status of the final star cluster.
Accordingly, the smaller SFE does not form a star cluster with a more rapidly declining $\lossig$ profile for the same embedded cluster model in MOND. In contrast, when the value of SFE decreases, the $\lossig$ profile declines more slowly with radius or even becomes flat after being evolved for a few Gyr. The observed $\lossig$ profile of NGC 2419 cannot be reproduced by reducing the value of SFE in MOND.
Although gas expulsion is not the sole mechanism for the observed kinematics in NGC 2419 in MOND, the slower decrease and the flat $\lossig$ profiles predicted by MOND are consistent with the existence of the observed flat $\lossig$ profiles for other GCs in the Galaxy \citep{Scarpa+2007,Scarpa+2011,Scarpa_Falomo2010,Lane+2009,Lane+2010,Hernandez_Jimenez2012,Hernandez+2013,Durazo+2017}. Although these GCs are not outer halo GCs and the external field from the Galaxy at their locations is about $1a_0$ \citep{Lane+2010}, the slowly declining and the flat $\sigma_{\rm LoS}$ profiles after gas expulsion for low SFEs are expected for them. Due to the fact that the external field enables the high energy stars to escape from the GCs, the strongly radially anisotropic orbits are lost. We will systematically study the tidal field and external field effects for GCs after undergoing gas expulsion in a future project.

\begin{figure}
\resizebox{8.5cm}{!}{\includegraphics{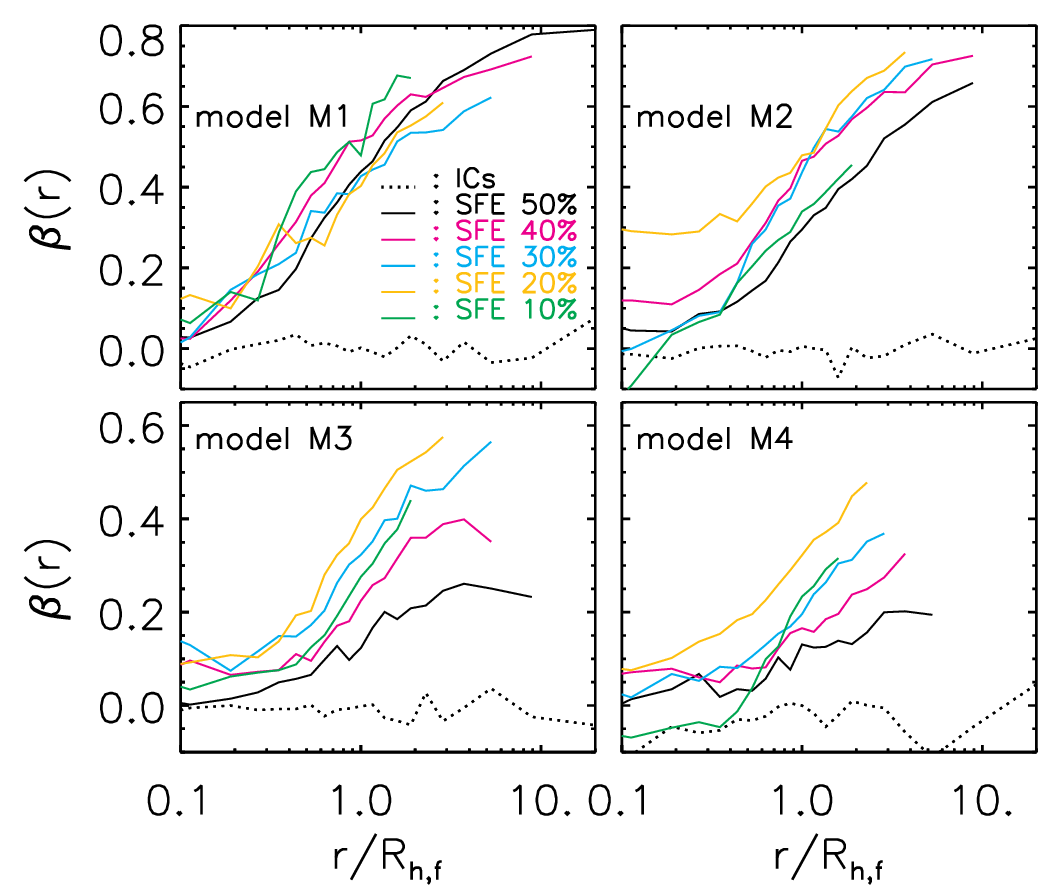}}
\caption{The anisotropy profiles, $\beta(r)$, of the ICs and the surviving star clusters evolved for $2\Gyr$ after gas expulsion. The radii are in units of $\rhf$ (and $\rh$ for the ICs). }
\label{beta}
\end{figure}

In general, the $\beta(r)$ profiles for all the surviving star clusters become radially anisotropic at the end of the simulations (Fig. \ref{beta}). The surviving star clusters are the most radially anisotropic for the massive and compact model, namely, model M$1$, with all values of SFE, while they are the least radially anisotropic for the least massive and diffuse model, model M$4$. 

For model M$1$, the $\beta(r)$ profiles are very similar for all the surviving star clusters with SFE between 0.1 and 0.5. The $\beta(r)$ profiles, starting from near $0$ in the central region, increase steeply to around $0.5$-$0.7$ at a radius of $2\rhf$, and then the profiles rise slowly out to large radii.
The values of $\beta$ at $R=\rhf$ for model M2 are between $0.4$ and $0.6$, for model M$3$ they are in the range of $[0.2,~0.5]$ and decrease to $0.1-0.4$ for model M$4$. Given that model M$1$ is dominated by quasi-Newtonian gravity, while model M$4$ is dominated by deep MOND, the radial anisotropy is weaker for models dominated by MOND gravity.
In a deep MOND system such as model M$4$, the internal gravity is much stronger than a pure Newtonian system with exactly the same density distribution. Therefore, the particles are more tightly bound to a deep MOND system. Since a MOND system does not expand as much as a Newtonian system when the two systems lose the same fraction of mass in gas, less radial orbits are generated in MOND after the gas expulsion.

\section{Conclusion and discussions}\label{summary}
In this work, we presented the first analysis of the line-of-sight velocity dispersion profiles and the anisotropy profiles of star clusters surviving gas expulsion in Newtonian and in Milgromian dynamics. A series of simulations with the SFE ranging from $10\%$-$50\%$ are performed for embedded clusters with initial masses in the range of $[10^5,10^7]\msun$. We summarise and discuss our main results here.

The LoS velocity dispersion of the surviving star clusters are studied in both Newtonian dynamics and MOND, with $\epsilon=0.5$ and $0.4$. In Newtonian dynamics, the $\lossig/\sigma_{\rm 0,f}$ profiles for the final star clusters are almost indistinguishable in the region where $R>\rhf$. In the inner region where $R<\rhf$, the models with smaller values of $\epsilon$ have larger values of $\sigma_{\rm LoS}/\sigma_{\rm 0,f}$. On the other hand, the value of the observed $\lossig$ profile for NGC2419 reaches $\approx 7\kms$ in the centre and drops to $1-2\kms$ at large radii where $R>3\re$ \citep[$\re$ is the effective radius of the star cluster,][]{Baumgardt+2009,Ibata+2011a,Ibata+2011b}. The existing dynamical models usually require a highly radially anisotropic profile for this star cluster \citep{Ibata+2011a,Ibata+2011b}. 
Here we have found that gas expulsion naturally provides an explanation for the observed $\lossig$ and the model-required $\beta(r)$ profiles of NGC 2419 in Newtonian dynamics. Furthermore, all the isolated Newtonian GCs should have a strongly radially anisotropic distribution of velocity dispersion at least in the outer regions of the star clusters ($R>\rh$), if gas expulsion with a low SFE is physically realistic for these GCs.

In contrast, GCs are allowed to have mildly or moderately radially anisotropic profiles in MOND undergoing gas expulsion with the same values of the SFE. In general, the central $\lossig/\sigma_{\rm 0,f}$ values are lower and the outer $\lossig/\sigma_{\rm 0,f}$ values are higher for the same model in MOND as in Newtonian dynamics. Radial-orbit instability occurs in the most massive MOND model with a SFE of 0.4, while the corresponding Newtonian model with the same model parameters is stable. The radial-orbit instability in the MOND post-gas-expulsion model comes from the bound high energy stars in the Milgromian gravitational potential. Such high energy stars escape from the final star cluster in Newtonian gravitation. Moreover, the $\lossig/\sigma_{\rm 0,f}$ profiles decline more slowly for the MOND models. The central $\sigma_{\rm LoS}/\sigma_{\rm 0,f}$ is smaller and the values of $\sigma_{\rm LoS}/\sigma_{\rm 0,f}$ in the outer regions are larger for the more diffuse models and also for models with smaller $\epsilon$. This stems from the gravitational potential well being deeper for the star clusters in MOND with the same density distribution as the Newtonian models, and the more diffuse models and the models with lower $\epsilon$ are in a deeper MOND regime.

In addition, star clusters can survive much lower $\epsilon_{\rm crit}$ in MOND \citep{Wu_Kroupa2017}. The $\lossig$ and $\beta(r)$ profiles are studied for the star clusters undergoing gas expulsion with $\epsilon=30\%$, $20\%$ and $10\%$. For the more compact embedded cluster, i.e., model M1, the $\lossig$ profile is almost flat for such a low SFE. When $\epsilon$ is down to $10\%$, the $\lossig$ profile is growing with radius. The possible reason is that the surviving star cluster is not fully virialised within the simulation time of $2~\Gyr$. When the model for the embedded cluster becomes more diffuse, the critical value of $\epsilon$ to generate a rapidly falling $\lossig$ profile decreases. For instance, the $\lossig$ profiles are falling fast with incrasing radius for $\epsilon\approx 0.2$ in models M3 and M4. In these models, only when $\epsilon$ is 0.1, the $\lossig$ profiles are flat. To summarise, the observed $\lossig$ profile of NGC 2419 cannot be reproduced by reducing the SFE undergoing gas expulsion in MOND.

The anisotropy profiles, $\beta(r)$, for these MOND models show that the more compact the embedded cluster is, the more radially anisotropic the final star cluster is. For the same embedded cluster with a compact density distribution (i.e., model M1), the $\beta(r)$ profiles for the final star clusters for different $\epsilon$ are very similar to each other. When the embedded cluster model becomes more diffuse, the final star clusters are more radially anisotropic for smaller values of $\epsilon$. Mildly ($\beta(r) \approx 0.2$ at large radii) and moderately ($\beta(r)<0.4$ at large radii) radial anisotropy profiles can be generated through gas expulsion in MOND, if the embedded cluster is a diffuse system and the gravity is dominated by deep MOND. This is because the particles in a MOND system are more tightly bound, and the gravitational potential does not change as much as that in quasi-Newtonian gravity after gas expulsion. There are fewer radial orbits generated by the sudden removal of gas. Therefore, MOND predicts a larger variety of velocity dispersion profiles and anisotropy profiles through the process of gas expulsion than Newtonian gravitation, which agrees with the variety of LoS velocity profiles for GCs observed in the Galactic halo.

We note in particular that while gas-expulsion-only models fail to account for the observed $\lossig$ profile of NGC 2419 (Fig. \ref{newtkin}) in MOND, if this cluster is on a very eccentric orbit then the time-varying external field effect will lead to a reproduction of the observed profile \citep{Wu_Kroupa2013}. At the same time, the gas expulsion process in MOND models can also account for flat outer $\lossig$ profiles, while the sole gas expulsion Newtonian models cannot. It has been suggested that GCs may reside in their own dark matter halos which would possibly account for the observed flat $\lossig$ profiles. Moreover, the possibility of existence of a dark matter halo around NGC 2419 is discussed in \citet{Ibata+2013}. The dynamical models which reproduce the surface mass density and kinematics of NGC 2419 cannot fully rule out a dark matter halo around this star cluster, but the dark matter particle hypothesis is challenged by other data \citep{Kroupa2012,Kroupa2015,Famaey_McGaugh2012}. It remains unknown whether this possibility is consistent with models of star cluster formation \citep{Moore1996}. Another physical mechanism is required in Newtonian dynamics to account for the observed flat $\lossig$ profiles.  

\section{Acknowledgments}
The authors thank the anonymous referee for the very valuable and helpful comments. The authors thank Luca Ciotti, Pasquale Londrillo and Carlo Nipoti for sharing their NMODY code. The authors acknowledge Ortwin Gerhard for sharing the code for generating the self-consistent N-body ICs using Lucy's method. XW thanks for support through grants from the Natural Science Foundation of China (Numbers NSFC-11503025 and NSFC-11421303), a Anhui Natural Science Foundation grant (Number 1708085MA20), and ``the Fundamental Research Funds for the Central Universities''. XW thanks for support from the ``Hundred Talents Project of Anhui Province''. 

\bibliographystyle{mn2e}
\bibliography{gasexp}

\begin{thebibliography}{}

\bibitem[\protect\citeauthoryear{{Aarseth}}{{Aarseth}}{2003}]{Aarseth2003}
{Aarseth} S.~J.,  2003, {Gravitational N-Body Simulations, by Sverre
  J.~Aarseth, pp.~430.~ISBN 0521432723.~Cambridge, UK: Cambridge University
  Press, November 2003.}

\bibitem[\protect\citeauthoryear{{Angus}}{{Angus}}{2009}]{Angus2009}
{Angus} G.~W.,  2009, \mnras, 394, 527

\bibitem[\protect\citeauthoryear{{Angus} \& {Diaferio}}{{Angus} \&
  {Diaferio}}{2011}]{Angus_Diaferio2011}
{Angus} G.~W.,  {Diaferio} A.,  2011, \mnras, 417, 941

\bibitem[\protect\citeauthoryear{{Angus}, {Diaferio}, {Famaey}, {Gentile} \&
  {van der Heyden}}{{Angus} et~al.}{2014}]{Angus+2014}
{Angus} G.~W.,  {Diaferio} A.,  {Famaey} B.,  {Gentile} G.,    {van der Heyden}
  K.~J.,  2014, JCAP, 10, 079

\bibitem[\protect\citeauthoryear{{Angus}, {Diaferio}, {Famaey} \& {van der
  Heyden}}{{Angus} et~al.}{2013}]{Angus+2013}
{Angus} G.~W.,  {Diaferio} A.,  {Famaey} B.,    {van der Heyden} K.~J.,  2013,
  \mnras, 436, 202

\bibitem[\protect\citeauthoryear{{Angus}, {Shan}, {Zhao} \& {Famaey}}{{Angus}
  et~al.}{2007}]{Angus+2007}
{Angus} G.~W.,  {Shan} H.~Y.,  {Zhao} H.~S.,    {Famaey} B.,  2007, \apjl, 654,
  L13

\bibitem[\protect\citeauthoryear{{Banerjee} \& {Kroupa}}{{Banerjee} \&
  {Kroupa}}{2012}]{Banerjee_Kroupa2012}
{Banerjee} S.,  {Kroupa} P.,  2012, \aap, 547, A23

\bibitem[\protect\citeauthoryear{{Banerjee} \& {Kroupa}}{{Banerjee} \&
  {Kroupa}}{2015}]{Banerjee_Kroupa2015}
{Banerjee} S.,  {Kroupa} P.,  2015, \mnras, 447, 728

\bibitem[\protect\citeauthoryear{{Banerjee} \& {Kroupa}}{{Banerjee} \&
  {Kroupa}}{2017}]{Banerjee_Kroupa2017}
{Banerjee} S.,  {Kroupa} P.,  2017, \aap, 597, A28

\bibitem[\protect\citeauthoryear{{Barnes}, {Williams}, {Babul} \&
  {Dalcanton}}{{Barnes} et~al.}{2005}]{Barnes+2005}
{Barnes} E.~I.,  {Williams} L.~L.~R.,  {Babul} A.,    {Dalcanton} J.~J.,  2005,
  \apj, 634, 775

\bibitem[\protect\citeauthoryear{{Bastian} \& {Goodwin}}{{Bastian} \&
  {Goodwin}}{2006}]{Bastian_Goodwin2006}
{Bastian} N.,  {Goodwin} S.~P.,  2006, \mnras, 369, L9

\bibitem[\protect\citeauthoryear{{Baumgardt}, {C{\^o}t{\'e}}, {Hilker},
  {Rejkuba}, {Mieske}, {Djorgovski} \& {Stetson}}{{Baumgardt}
  et~al.}{2009}]{Baumgardt+2009}
{Baumgardt} H.,  {C{\^o}t{\'e}} P.,  {Hilker} M.,  {Rejkuba} M.,  {Mieske} S.,
  {Djorgovski} S.~G.,    {Stetson} P.,  2009, \mnras, 396, 2051

\bibitem[\protect\citeauthoryear{{Baumgardt}, {Grebel} \& {Kroupa}}{{Baumgardt}
  et~al.}{2005}]{Baumgardt+2005}
{Baumgardt} H.,  {Grebel} E.~K.,    {Kroupa} P.,  2005, \mnras, 359, L1

\bibitem[\protect\citeauthoryear{{Baumgardt} \& {Kroupa}}{{Baumgardt} \&
  {Kroupa}}{2007}]{Baumgardt_Kroupa2007}
{Baumgardt} H.,  {Kroupa} P.,  2007, \mnras, 380, 1589

\bibitem[\protect\citeauthoryear{{Baumgardt}, {Makino} \& {Hut}}{{Baumgardt}
  et~al.}{2005}]{Baumgardt+2005b}
{Baumgardt} H.,  {Makino} J.,    {Hut} P.,  2005, \apj, 620, 238

\bibitem[\protect\citeauthoryear{{Bekenstein} \& {Milgrom}}{{Bekenstein} \&
  {Milgrom}}{1984}]{BM1984}
{Bekenstein} J.,  {Milgrom} M.,  1984, \apj, 286, 7

\bibitem[\protect\citeauthoryear{{Bellazzini}}{{Bellazzini}}{2007}]{Bellazzini%
2007}
{Bellazzini} M.,  2007, \aap, 473, 171

\bibitem[\protect\citeauthoryear{{Bellazzini}, {Dalessandro}, {Sollima} \&
  {Ibata}}{{Bellazzini} et~al.}{2012}]{Bellazzini+2012}
{Bellazzini} M.,  {Dalessandro} E.,  {Sollima} A.,    {Ibata} R.,  2012,
  \mnras, 423, 844

\bibitem[\protect\citeauthoryear{{Bellovary}, {Dalcanton}, {Babul}, {Quinn},
  {Maas}, {Austin}, {Williams} \& {Barnes}}{{Bellovary}
  et~al.}{2008}]{Bellovary+2008}
{Bellovary} J.~M.,  {Dalcanton} J.~J.,  {Babul} A.,  {Quinn} T.~R.,  {Maas}
  R.~W.,  {Austin} C.~G.,  {Williams} L.~L.~R.,    {Barnes} E.~I.,  2008, \apj,
  685, 739

\bibitem[\protect\citeauthoryear{{Bertin}, {Pegoraro}, {Rubini} \&
  {Vesperini}}{{Bertin} et~al.}{1994}]{Bertin+1994}
{Bertin} G.,  {Pegoraro} F.,  {Rubini} F.,    {Vesperini} E.,  1994, \apj, 434,
  94

\bibitem[\protect\citeauthoryear{{Bertin} \& {Stiavelli}}{{Bertin} \&
  {Stiavelli}}{1989}]{Bertin_Stiavelli1989}
{Bertin} G.,  {Stiavelli} M.,  1989, \apj, 338, 723

\bibitem[\protect\citeauthoryear{{Bertin} \& {Trenti}}{{Bertin} \&
  {Trenti}}{2003}]{Bertin_Trenti2003}
{Bertin} G.,  {Trenti} M.,  2003, \apj, 584, 729

\bibitem[\protect\citeauthoryear{{Boily} \& {Kroupa}}{{Boily} \&
  {Kroupa}}{2003}]{Boily_Kroupa2003}
{Boily} C.~M.,  {Kroupa} P.,  2003, \mnras, 338, 673

\bibitem[\protect\citeauthoryear{{Brinkmann}, {Banerjee}, {Motwani} \&
  {Kroupa}}{{Brinkmann} et~al.}{2017}]{Brinkmann+2017}
{Brinkmann} N.,  {Banerjee} S.,  {Motwani} B.,    {Kroupa} P.,  2017, \aap,
  600, A49

\bibitem[\protect\citeauthoryear{{Br{\"u}ns} \& {Kroupa}}{{Br{\"u}ns} \&
  {Kroupa}}{2011}]{Bruens_Kroupa2011}
{Br{\"u}ns} R.~C.,  {Kroupa} P.,  2011, \apj, 729, 69

\bibitem[\protect\citeauthoryear{{Casetti-Dinescu}, {Girard}, {Majewski},
  {Vivas}, {Wilhelm}, {Carlin}, {Beers} \& {van Altena}}{{Casetti-Dinescu}
  et~al.}{2009}]{Casetti-Dinescu+2009}
{Casetti-Dinescu} D.~I.,  {Girard} T.~M.,  {Majewski} S.~R.,  {Vivas} A.~K.,
  {Wilhelm} R.,  {Carlin} J.~L.,  {Beers} T.~C.,    {van Altena} W.~F.,  2009,
  \apjl, 701, L29

\bibitem[\protect\citeauthoryear{{Clowe}, {Brada{\v c}}, {Gonzalez},
  {Markevitch}, {Randall}, {Jones} \& {Zaritsky}}{{Clowe}
  et~al.}{2006}]{Clowe+2006}
{Clowe} D.,  {Brada{\v c}} M.,  {Gonzalez} A.~H.,  {Markevitch} M.,  {Randall}
  S.~W.,  {Jones} C.,    {Zaritsky} D.,  2006, \apjl, 648, L109

\bibitem[\protect\citeauthoryear{{Conroy}, {Loeb} \& {Spergel}}{{Conroy}
  et~al.}{2011}]{Conroy+2011}
{Conroy} C.,  {Loeb} A.,    {Spergel} D.~N.,  2011, \apj, 741, 72

\bibitem[\protect\citeauthoryear{{Derakhshani} \& {Haghi}}{{Derakhshani} \&
  {Haghi}}{2014}]{Derakhshani_Haghi2014}
{Derakhshani} K.,  {Haghi} H.,  2014, \apj, 785, 166

\bibitem[\protect\citeauthoryear{Di~Cintio \& Lelli}{Di~Cintio \&
  Lelli}{2016}]{DiCintio+2016}
Di~Cintio A.,  Lelli F.,  2016, \mnras: Letters, 456, L127

\bibitem[\protect\citeauthoryear{{Di Criscienzo}, {Greco}, {Ripepi},
  {Clementini}, {Dall'Ora}, {Marconi}, {Musella}, {Federici} \& {Di
  Fabrizio}}{{Di Criscienzo} et~al.}{2011}]{DiCriscienzo+2011}
{Di Criscienzo} M.,  {Greco} C.,  {Ripepi} V.,  {Clementini} G.,  {Dall'Ora}
  M.,  {Marconi} M.,  {Musella} I.,  {Federici} L.,    {Di Fabrizio} L.,  2011,
  \aj, 141, 81

\bibitem[\protect\citeauthoryear{{Dib}, {Gutkin}, {Brandner} \& {Basu}}{{Dib}
  et~al.}{2013}]{Dib+2013}
{Dib} S.,  {Gutkin} J.,  {Brandner} W.,    {Basu} S.,  2013, \mnras, 436, 3727

\bibitem[\protect\citeauthoryear{{Durazo}, {Hernandez}, {Cervantes Sodi} \&
  {S{\'a}nchez}}{{Durazo} et~al.}{2017}]{Durazo+2017}
{Durazo} R.,  {Hernandez} X.,  {Cervantes Sodi} B.,    {S{\'a}nchez} S.~F.,
  2017, \apj, 837, 179

\bibitem[\protect\citeauthoryear{{Famaey} \& {Binney}}{{Famaey} \&
  {Binney}}{2005}]{Famaey_Binney2005}
{Famaey} B.,  {Binney} J.,  2005, \mnras, 363, 603

\bibitem[\protect\citeauthoryear{{Famaey}, {Bruneton} \& {Zhao}}{{Famaey}
  et~al.}{2007}]{Famaey+2007}
{Famaey} B.,  {Bruneton} J.-P.,    {Zhao} H.,  2007, \mnras, 377, L79

\bibitem[\protect\citeauthoryear{{Famaey} \& {McGaugh}}{{Famaey} \&
  {McGaugh}}{2012}]{Famaey_McGaugh2012}
{Famaey} B.,  {McGaugh} S.~S.,  2012, Living Reviews in Relativity, 15, 10

\bibitem[\protect\citeauthoryear{{Fellhauer} \& {Kroupa}}{{Fellhauer} \&
  {Kroupa}}{2005}]{Fellhauer_Kroupa2005}
{Fellhauer} M.,  {Kroupa} P.,  2005, \apj, 630, 879

\bibitem[\protect\citeauthoryear{{Geyer} \& {Burkert}}{{Geyer} \&
  {Burkert}}{2001}]{Geyer_Burkert2001}
{Geyer} M.~P.,  {Burkert} A.,  2001, in {Funes} J.~G.,  {Corsini} E.~M.,  eds,
  Galaxy Disks and Disk Galaxies Vol.~230 of Astronomical Society of the
  Pacific Conference Series, {Gas Expulsion from Star Forming Regions and the
  Formation of Globular Clusters}.
pp 311--312

\bibitem[\protect\citeauthoryear{{Goodwin}}{{Goodwin}}{1997}]{Goodwin1997}
{Goodwin} S.~P.,  1997, \mnras, 284, 785

\bibitem[\protect\citeauthoryear{{Goodwin} \& {Bastian}}{{Goodwin} \&
  {Bastian}}{2006}]{Goodwin_bastian2006}
{Goodwin} S.~P.,  {Bastian} N.,  2006, \mnras, 373, 752

\bibitem[\protect\citeauthoryear{{Haghi}, {Baumgardt}, {Kroupa}, {Grebel},
  {Hilker} \& {Jordi}}{{Haghi} et~al.}{2009}]{Haghi+2009}
{Haghi} H.,  {Baumgardt} H.,  {Kroupa} P.,  {Grebel} E.~K.,  {Hilker} M.,
  {Jordi} K.,  2009, \mnras, 395, 1549

\bibitem[\protect\citeauthoryear{{Haghi}, {Zonoozi}, {Kroupa}, {Banerjee} \&
  {Baumgardt}}{{Haghi} et~al.}{2015}]{Haghi+2015}
{Haghi} H.,  {Zonoozi} A.~H.,  {Kroupa} P.,  {Banerjee} S.,    {Baumgardt} H.,
  2015, \mnras, 454, 3872

\bibitem[\protect\citeauthoryear{{Hansen}, {Klein}, {McKee} \&
  {Fisher}}{{Hansen} et~al.}{2012}]{Hansen+2012}
{Hansen} C.~E.,  {Klein} R.~I.,  {McKee} C.~F.,    {Fisher} R.~T.,  2012, \apj,
  747, 22

\bibitem[\protect\citeauthoryear{{Harris}}{{Harris}}{1996}]{Harris1996}
{Harris} W.~E.,  1996, \aj, 112, 1487

\bibitem[\protect\citeauthoryear{{Hernandez} \& {Jim{\'e}nez}}{{Hernandez} \&
  {Jim{\'e}nez}}{2012}]{Hernandez_Jimenez2012}
{Hernandez} X.,  {Jim{\'e}nez} M.~A.,  2012, \apj, 750, 9

\bibitem[\protect\citeauthoryear{{Hernandez}, {Jim{\'e}nez} \&
  {Allen}}{{Hernandez} et~al.}{2013}]{Hernandez+2013}
{Hernandez} X.,  {Jim{\'e}nez} M.~A.,    {Allen} C.,  2013, \mnras, 428, 3196

\bibitem[\protect\citeauthoryear{{Higuchi}, {Kurono}, {Saito} \&
  {Kawabe}}{{Higuchi} et~al.}{2009}]{Higuchi+2009}
{Higuchi} A.~E.,  {Kurono} Y.,  {Saito} M.,    {Kawabe} R.,  2009, \apj, 705,
  468

\bibitem[\protect\citeauthoryear{{Hopkins}, {Narayanan}, {Murray} \&
  {Quataert}}{{Hopkins} et~al.}{2013}]{Hopkins+2013}
{Hopkins} P.~F.,  {Narayanan} D.,  {Murray} N.,    {Quataert} E.,  2013,
  \mnras, 433, 69

\bibitem[\protect\citeauthoryear{{Ibata}, {Nipoti}, {Sollima}, {Bellazzini},
  {Chapman} \& {Dalessandro}}{{Ibata} et~al.}{2013}]{Ibata+2013}
{Ibata} R.,  {Nipoti} C.,  {Sollima} A.,  {Bellazzini} M.,  {Chapman} S.~C.,
  {Dalessandro} E.,  2013, \mnras, 428, 3648

\bibitem[\protect\citeauthoryear{{Ibata}, {Sollima}, {Nipoti}, {Bellazzini},
  {Chapman} \& {Dalessandro}}{{Ibata} et~al.}{2011a}]{Ibata+2011b}
{Ibata} R.,  {Sollima} A.,  {Nipoti} C.,  {Bellazzini} M.,  {Chapman} S.~C.,
  {Dalessandro} E.,  2011a, \apj, 743, 43

\bibitem[\protect\citeauthoryear{{Ibata}, {Sollima}, {Nipoti}, {Bellazzini},
  {Chapman} \& {Dalessandro}}{{Ibata} et~al.}{2011b}]{Ibata+2011a}
{Ibata} R.,  {Sollima} A.,  {Nipoti} C.,  {Bellazzini} M.,  {Chapman} S.~C.,
  {Dalessandro} E.,  2011b, \apj, 738, 186

\bibitem[\protect\citeauthoryear{{Jee}, {Mahdavi}, {Hoekstra}, {Babul},
  {Dalcanton}, {Carroll} \& {Capak}}{{Jee} et~al.}{2012}]{Jee+2012}
{Jee} M.~J.,  {Mahdavi} A.,  {Hoekstra} H.,  {Babul} A.,  {Dalcanton} J.~J.,
  {Carroll} P.,    {Capak} P.,  2012, \apj, 747, 96

\bibitem[\protect\citeauthoryear{{Kainulainen}, {Federrath} \&
  {Henning}}{{Kainulainen} et~al.}{2014}]{Kainulainen+2014}
{Kainulainen} J.,  {Federrath} C.,    {Henning} T.,  2014, Science, 344, 183

\bibitem[\protect\citeauthoryear{Keller \& Wadsley}{Keller \&
  Wadsley}{2017}]{Keller_Wadsley2017}
Keller B.~W.,  Wadsley J.~W.,  2017, The Astrophysical Journal Letters, 835,
  L17

\bibitem[\protect\citeauthoryear{{Kn{\"o}dlseder}}{{Kn{\"o}dlseder}}{2000}]{Kn%
odlseder2000}
{Kn{\"o}dlseder} J.,  2000, \aap, 360, 539

\bibitem[\protect\citeauthoryear{{Kroupa}}{{Kroupa}}{2005}]{Kroupa2005}
{Kroupa} P.,  2005, in {Turon} C.,  {O'Flaherty} K.~S.,   {Perryman} M.~A.~C.,
  eds, The Three-Dimensional Universe with Gaia Vol.~576 of ESA Special
  Publication, {The Fundamental Building Blocks of Galaxies}.
p.~629

\bibitem[\protect\citeauthoryear{{Kroupa}}{{Kroupa}}{2008}]{Kroupa2008}
{Kroupa} P.,  2008, in {Aarseth} S.~J.,  {Tout} C.~A.,   {Mardling} R.~A.,
  eds, The Cambridge N-Body Lectures Vol.~760 of Lecture Notes in Physics,
  Berlin Springer Verlag, {Initial Conditions for Star Clusters}.
p.~181

\bibitem[\protect\citeauthoryear{{Kroupa}}{{Kroupa}}{2012}]{Kroupa2012}
{Kroupa} P.,  2012, {PASA}, 29, 395

\bibitem[\protect\citeauthoryear{{Kroupa}}{{Kroupa}}{2015}]{Kroupa2015}
{Kroupa} P.,  2015, Canadian Journal of Physics, 93, 169

\bibitem[\protect\citeauthoryear{{Kroupa}, {Aarseth} \& {Hurley}}{{Kroupa}
  et~al.}{2001}]{Kroupa+2001}
{Kroupa} P.,  {Aarseth} S.,    {Hurley} J.,  2001, \mnras, 321, 699

\bibitem[\protect\citeauthoryear{{Kroupa}, {Famaey}, {de Boer},
  {Dabringhausen}, {Pawlowski}, {Boily}, {Jerjen}, {Forbes}, {Hensler} \&
  {Metz}}{{Kroupa} et~al.}{2010}]{Kroupa+2010}
{Kroupa} P.,  {Famaey} B.,  {de Boer} K.~S.,  {Dabringhausen} J.,  {Pawlowski}
  M.~S.,  {Boily} C.~M.,  {Jerjen} H.,  {Forbes} D.,  {Hensler} G.,    {Metz}
  M.,  2010, \aap, 523, A32

\bibitem[\protect\citeauthoryear{{Kroupa}, {Haghi}, {Javanmardi}, {Zonoozi},
  {M{\"u}ller}, {Banik}, {Wu}, {Zhao} \& {Dabringhausen}}{{Kroupa}
  et~al.}{2018}]{Kroupa+2018nature}
{Kroupa} P.,  {Haghi} H.,  {Javanmardi} B.,  {Zonoozi} A.~H.,  {M{\"u}ller} O.,
   {Banik} I.,  {Wu} X.,  {Zhao} H.,    {Dabringhausen} J.,  2018, \nat, 561,
  E4

\bibitem[\protect\citeauthoryear{{Kroupa}, {Je{\v r}{\'a}bkov{\'a}},
  {Dinnbier}, {Beccari} \& {Yan}}{{Kroupa} et~al.}{2018}]{Kroupa+2018}
{Kroupa} P.,  {Je{\v r}{\'a}bkov{\'a}} T.,  {Dinnbier} F.,  {Beccari} G.,
  {Yan} Z.,  2018, \aap, 612, A74

\bibitem[\protect\citeauthoryear{{Krumholz} \& {Matzner}}{{Krumholz} \&
  {Matzner}}{2009}]{Krumholz_Matzner2009}
{Krumholz} M.~R.,  {Matzner} C.~D.,  2009, \apj, 703, 1352

\bibitem[\protect\citeauthoryear{{Lada} \& {Lada}}{{Lada} \&
  {Lada}}{2003}]{Lada_Lada2003}
{Lada} C.~J.,  {Lada} E.~A.,  2003, \araa, 41, 57

\bibitem[\protect\citeauthoryear{{Lada}, {Margulis} \& {Dearborn}}{{Lada}
  et~al.}{1984}]{Lada+1984}
{Lada} C.~J.,  {Margulis} M.,    {Dearborn} D.,  1984, \apj, 285, 141

\bibitem[\protect\citeauthoryear{{Lane}, {Kiss}, {Lewis}, {Ibata}, {Siebert},
  {Bedding} \& {Sz{\'e}kely}}{{Lane} et~al.}{2009}]{Lane+2009}
{Lane} R.~R.,  {Kiss} L.~L.,  {Lewis} G.~F.,  {Ibata} R.~A.,  {Siebert} A.,
  {Bedding} T.~R.,    {Sz{\'e}kely} P.,  2009, \mnras, 400, 917

\bibitem[\protect\citeauthoryear{{Lane}, {Kiss}, {Lewis}, {Ibata}, {Siebert},
  {Bedding}, {Sz{\'e}kely}, {Balog} \& {Szab{\'o}}}{{Lane}
  et~al.}{2010}]{Lane+2010}
{Lane} R.~R.,  {Kiss} L.~L.,  {Lewis} G.~F.,  {Ibata} R.~A.,  {Siebert} A.,
  {Bedding} T.~R.,  {Sz{\'e}kely} P.,  {Balog} Z.,    {Szab{\'o}} G.~M.,  2010,
  \mnras, 406, 2732

\bibitem[\protect\citeauthoryear{{Londrillo} \& {Nipoti}}{{Londrillo} \&
  {Nipoti}}{2009}]{nmody}
{Londrillo} P.,  {Nipoti} C.,  2009, Memorie della Societa Astronomica Italiana
  Supplement, 13, 89

\bibitem[\protect\citeauthoryear{{Lucy}}{{Lucy}}{1974}]{Lucy1974}
{Lucy} L.~B.,  1974, \aj, 79, 745

\bibitem[\protect\citeauthoryear{Ludlow, Ben\'{\i}tez-Llambay, Schaller,
  Theuns, Frenk, Bower, Schaye, Crain, Navarro, Fattahi \& Oman}{Ludlow
  et~al.}{2017}]{Ludlow+2017}
Ludlow A.~D.,  Ben\'{\i}tez-Llambay A.,  Schaller M.,  Theuns T.,  Frenk C.~S.,
   Bower R.,  Schaye J.,  Crain R.~A.,  Navarro J.~F.,  Fattahi A.,    Oman
  K.~A.,  2017, Phys. Rev. Lett., 118, 161103

\bibitem[\protect\citeauthoryear{{Lynden-Bell}}{{Lynden-Bell}}{1967}]{Lynden-B%
ell1967}
{Lynden-Bell} D.,  1967, \mnras, 136, 101

\bibitem[\protect\citeauthoryear{{Mackey} \& {van den Bergh}}{{Mackey} \& {van
  den Bergh}}{2005}]{Mackey_vandenBergh2005}
{Mackey} A.~D.,  {van den Bergh} S.,  2005, \mnras, 360, 631

\bibitem[\protect\citeauthoryear{{Madejsky} \& {Bien}}{{Madejsky} \&
  {Bien}}{1993}]{Madejsky_Bien1993}
{Madejsky} R.,  {Bien} R.,  1993, \aap, 280, 383

\bibitem[\protect\citeauthoryear{{Marks} \& {Kroupa}}{{Marks} \&
  {Kroupa}}{2011}]{Marks_Kroupa2011}
{Marks} M.,  {Kroupa} P.,  2011, \mnras, 417, 1702

\bibitem[\protect\citeauthoryear{{Marks}, {Kroupa}, {Dabringhausen} \&
  {Pawlowski}}{{Marks} et~al.}{2012}]{Marks+2012}
{Marks} M.,  {Kroupa} P.,  {Dabringhausen} J.,    {Pawlowski} M.~S.,  2012,
  \mnras, 422, 2246

\bibitem[\protect\citeauthoryear{{May} \& {Binney}}{{May} \&
  {Binney}}{1986}]{May_Binney1986}
{May} A.,  {Binney} J.,  1986, \mnras, 221, 13P

\bibitem[\protect\citeauthoryear{{McGaugh}}{{McGaugh}}{2004}]{McGaugh2004}
{McGaugh} S.~S.,  2004, \apj, 609, 652

\bibitem[\protect\citeauthoryear{{McGaugh}}{{McGaugh}}{2012}]{McGaugh2012}
{McGaugh} S.~S.,  2012, \aj, 143, 40

\bibitem[\protect\citeauthoryear{{Megeath}, {Gutermuth}, {Muzerolle},
  {Kryukova}, {Hora}, {Allen}, {Flaherty}, {Hartmann}, {Myers}, {Pipher},
  {Stauffer}, {Young} \& {Fazio}}{{Megeath} et~al.}{2016}]{Megeath+2016}
{Megeath} S.~T.,  {Gutermuth} R.,  {Muzerolle} J.,  {Kryukova} E.,  {Hora}
  J.~L.,  {Allen} L.~E.,  {Flaherty} K.,  {Hartmann} L.,  {Myers} P.~C.,
  {Pipher} J.~L.,  {Stauffer} J.,  {Young} E.~T.,    {Fazio} G.~G.,  2016, \aj,
  151, 5

\bibitem[\protect\citeauthoryear{{Milgrom}}{{Milgrom}}{1983}]{Milgrom1983a}
{Milgrom} M.,  1983, \apj, 270, 365

\bibitem[\protect\citeauthoryear{{Milgrom}}{{Milgrom}}{1986}]{Milgrom1986}
{Milgrom} M.,  1986, \apj, 306, 9

\bibitem[\protect\citeauthoryear{{Milgrom}}{{Milgrom}}{1994}]{Milgrom1994}
{Milgrom} M.,  1994, \apj, 429, 540

\bibitem[\protect\citeauthoryear{{Moore}}{{Moore}}{1996}]{Moore1996}
{Moore} B.,  1996, \apjl, 461, L13

\bibitem[\protect\citeauthoryear{{Murray}}{{Murray}}{2011}]{Murray2011}
{Murray} N.,  2011, \apj, 729, 133

\bibitem[\protect\citeauthoryear{{Murray}, {Quataert} \& {Thompson}}{{Murray}
  et~al.}{2010}]{Murray+2010}
{Murray} N.,  {Quataert} E.,    {Thompson} T.~A.,  2010, \apj, 709, 191

\bibitem[\protect\citeauthoryear{{Natarajan} \& {Zhao}}{{Natarajan} \&
  {Zhao}}{2008}]{Natarajan_Zhao2008}
{Natarajan} P.,  {Zhao} H.,  2008, \mnras, 389, 250

\bibitem[\protect\citeauthoryear{Navarro, Benítez-Llambay, Fattahi, Frenk,
  Ludlow, Oman, Schaller \& Theuns}{Navarro et~al.}{2017}]{Navarro+2017}
Navarro J.~F.,  Benítez-Llambay A.,  Fattahi A.,  Frenk C.~S.,  Ludlow A.~D.,
  Oman K.~A.,  Schaller M.,    Theuns T.,  2017, \mnras, 471, 1841

\bibitem[\protect\citeauthoryear{{Nipoti}, {Ciotti}, {Binney} \&
  {Londrillo}}{{Nipoti} et~al.}{2008}]{Nipoti+2008}
{Nipoti} C.,  {Ciotti} L.,  {Binney} J.,    {Londrillo} P.,  2008, \mnras, 386,
  2194

\bibitem[\protect\citeauthoryear{{Nipoti}, {Ciotti} \& {Londrillo}}{{Nipoti}
  et~al.}{2011}]{Nipoti+2011}
{Nipoti} C.,  {Ciotti} L.,    {Londrillo} P.,  2011, \mnras, 414, 3298

\bibitem[\protect\citeauthoryear{{Nipoti}, {Londrillo} \& {Ciotti}}{{Nipoti}
  et~al.}{2007}]{Nipoti+2007}
{Nipoti} C.,  {Londrillo} P.,    {Ciotti} L.,  2007, \apj, 660, 256

\bibitem[\protect\citeauthoryear{{Pfalzner} \& {Kaczmarek}}{{Pfalzner} \&
  {Kaczmarek}}{2013}]{Pfalzner_Kaczmarek2013}
{Pfalzner} S.,  {Kaczmarek} T.,  2013, \aap, 555, A135

\bibitem[\protect\citeauthoryear{{Plummer}}{{Plummer}}{1911}]{Plummer1911}
{Plummer} H.~C.,  1911, \mnras, 71, 460

\bibitem[\protect\citeauthoryear{{Polyachenko} \& {Shukhman}}{{Polyachenko} \&
  {Shukhman}}{1981}]{Polyachenko_Shukhman1981}
{Polyachenko} V.~L.,  {Shukhman} I.~G.,  1981, \sovast, 25, 533

\bibitem[\protect\citeauthoryear{Read, Iorio, Agertz \& Fraternali}{Read
  et~al.}{2016}]{Read+2016}
Read J.~I.,  Iorio G.,  Agertz O.,    Fraternali F.,  2016, \mnras, 462, 3628

\bibitem[\protect\citeauthoryear{{Saha}}{{Saha}}{1991}]{Saha1991}
{Saha} P.,  1991, \mnras, 248, 494

\bibitem[\protect\citeauthoryear{{Saha}}{{Saha}}{1992}]{Saha1992}
{Saha} P.,  1992, \mnras, 254, 132

\bibitem[\protect\citeauthoryear{{Sanders}}{{Sanders}}{1990}]{Sanders1990}
{Sanders} R.~H.,  1990, \aapr, 2, 1

\bibitem[\protect\citeauthoryear{{Sanders}}{{Sanders}}{2012a}]{Sanders2012b}
{Sanders} R.~H.,  2012a, \mnras, 419, L6

\bibitem[\protect\citeauthoryear{{Sanders}}{{Sanders}}{2012b}]{Sanders2012a}
{Sanders} R.~H.,  2012b, \mnras, 422, L21

\bibitem[\protect\citeauthoryear{{Sanders} \& {McGaugh}}{{Sanders} \&
  {McGaugh}}{2002}]{Sanders_McGaugh2002}
{Sanders} R.~H.,  {McGaugh} S.~S.,  2002, \araa, 40, 263

\bibitem[\protect\citeauthoryear{{Scarpa} \& {Falomo}}{{Scarpa} \&
  {Falomo}}{2010}]{Scarpa_Falomo2010}
{Scarpa} R.,  {Falomo} R.,  2010, \aap, 523, A43

\bibitem[\protect\citeauthoryear{{Scarpa}, {Marconi}, {Carraro}, {Falomo} \&
  {Villanova}}{{Scarpa} et~al.}{2011}]{Scarpa+2011}
{Scarpa} R.,  {Marconi} G.,  {Carraro} G.,  {Falomo} R.,    {Villanova} S.,
  2011, \aap, 525, A148

\bibitem[\protect\citeauthoryear{{Scarpa}, {Marconi}, {Gilmozzi} \&
  {Carraro}}{{Scarpa} et~al.}{2007}]{Scarpa+2007}
{Scarpa} R.,  {Marconi} G.,  {Gilmozzi} R.,    {Carraro} G.,  2007, \aap, 462,
  L9

\bibitem[\protect\citeauthoryear{{Shetrone}, {C{\^o}t{\'e}} \&
  {Sargent}}{{Shetrone} et~al.}{2001}]{Shetrone+2001}
{Shetrone} M.~D.,  {C{\^o}t{\'e}} P.,    {Sargent} W.~L.~W.,  2001, \apj, 548,
  592

\bibitem[\protect\citeauthoryear{{Skordis}, {Mota}, {Ferreira} \&
  {B{\oe}hm}}{{Skordis} et~al.}{2006}]{Skordis+2006}
{Skordis} C.,  {Mota} D.~F.,  {Ferreira} P.~G.,    {B{\oe}hm} C.,  2006,
  Physical Review Letters, 96, 011301

\bibitem[\protect\citeauthoryear{{Smith}, {Stassun} \& {Bally}}{{Smith}
  et~al.}{2005}]{Smith+2005}
{Smith} N.,  {Stassun} K.~G.,    {Bally} J.,  2005, \aj, 129, 888

\bibitem[\protect\citeauthoryear{{Sollima} \& {Nipoti}}{{Sollima} \&
  {Nipoti}}{2010}]{Sollima_Nipoti2010}
{Sollima} A.,  {Nipoti} C.,  2010, \mnras, 401, 131

\bibitem[\protect\citeauthoryear{{Tilley} \& {Pudritz}}{{Tilley} \&
  {Pudritz}}{2004}]{Tilley_Pudritz2004}
{Tilley} D.~A.,  {Pudritz} R.~E.,  2004, \mnras, 353, 769

\bibitem[\protect\citeauthoryear{{Trager}, {King} \& {Djorgovski}}{{Trager}
  et~al.}{1995}]{Trager+1995}
{Trager} S.~C.,  {King} I.~R.,    {Djorgovski} S.,  1995, \aj, 109, 218

\bibitem[\protect\citeauthoryear{{Trenti} \& {Bertin}}{{Trenti} \&
  {Bertin}}{2006}]{Trenti_Bertin2006}
{Trenti} M.,  {Bertin} G.,  2006, \apj, 637, 717

\bibitem[\protect\citeauthoryear{{Whitmore}, {Zhang}, {Leitherer}, {Fall},
  {Schweizer} \& {Miller}}{{Whitmore} et~al.}{1999}]{Whitmore+1999}
{Whitmore} B.~C.,  {Zhang} Q.,  {Leitherer} C.,  {Fall} S.~M.,  {Schweizer} F.,
     {Miller} B.~W.,  1999, \aj, 118, 1551

\bibitem[\protect\citeauthoryear{{Wu}, {Famaey}, {Gentile}, {Perets} \&
  {Zhao}}{{Wu} et~al.}{2008}]{Wu+2008}
{Wu} X.,  {Famaey} B.,  {Gentile} G.,  {Perets} H.,    {Zhao} H.,  2008,
  \mnras, 386, 2199

\bibitem[\protect\citeauthoryear{{Wu} \& {Kroupa}}{{Wu} \&
  {Kroupa}}{2013}]{Wu_Kroupa2013}
{Wu} X.,  {Kroupa} P.,  2013, \mnras, 435, 728

\bibitem[\protect\citeauthoryear{{Wu} \& {Kroupa}}{{Wu} \&
  {Kroupa}}{2015}]{Wu_Kroupa2015}
{Wu} X.,  {Kroupa} P.,  2015, \mnras, 446, 330

\bibitem[\protect\citeauthoryear{{Wu} \& {Kroupa}}{{Wu} \&
  {Kroupa}}{2018}]{Wu_Kroupa2017}
{Wu} X.,  {Kroupa} P.,  2018, \apj, 853, 60

\bibitem[\protect\citeauthoryear{{Wu}, {Wang}, {Feix} \& {Zhao}}{{Wu}
  et~al.}{2017}]{Wu+2017}
{Wu} X.,  {Wang} Y.,  {Feix} M.,    {Zhao} H.,  2017, \apj, 844, 130

\bibitem[\protect\citeauthoryear{{Zhang}, {Fall} \& {Whitmore}}{{Zhang}
  et~al.}{2001}]{Zhang+2001}
{Zhang} Q.,  {Fall} S.~M.,    {Whitmore} B.~C.,  2001, \apj, 561, 727

\bibitem[\protect\citeauthoryear{{Zhao}}{{Zhao}}{2008}]{Zhao2008}
{Zhao} H.~S.,  2008, in Journal of Physics Conference Series Vol.~140 of
  Journal of Physics Conference Series, {An ecological approach to problems of
  Dark Energy, Dark Matter, MOND and Neutrinos}.
p. 012002

\bibitem[\protect\citeauthoryear{{Zocchi}, {Bertin} \& {Varri}}{{Zocchi}
  et~al.}{2012}]{Zocchi+2012}
{Zocchi} A.,  {Bertin} G.,    {Varri} A.~L.,  2012, \aap, 539, A65

\bibitem[\protect\citeauthoryear{{Zonoozi}, {Haghi}, {Kroupa}, {K{\"u}pper} \&
  {Baumgardt}}{{Zonoozi} et~al.}{2017}]{Zonoozi+2017}
{Zonoozi} A.~H.,  {Haghi} H.,  {Kroupa} P.,  {K{\"u}pper} A.~H.~W.,
  {Baumgardt} H.,  2017, \mnras, 467, 758

\bibitem[\protect\citeauthoryear{{Zonoozi}, {Haghi}, {K{\"u}pper}, {Baumgardt},
  {Frank} \& {Kroupa}}{{Zonoozi} et~al.}{2014}]{Zonoozi+2014}
{Zonoozi} A.~H.,  {Haghi} H.,  {K{\"u}pper} A.~H.~W.,  {Baumgardt} H.,  {Frank}
  M.~J.,    {Kroupa} P.,  2014, \mnras, 440, 3172

\bibitem[\protect\citeauthoryear{{Zonoozi}, {K{\"u}pper}, {Baumgardt}, {Haghi},
  {Kroupa} \& {Hilker}}{{Zonoozi} et~al.}{2011}]{Zonoozi+2011}
{Zonoozi} A.~H.,  {K{\"u}pper} A.~H.~W.,  {Baumgardt} H.,  {Haghi} H.,
  {Kroupa} P.,    {Hilker} M.,  2011, \mnras, 411, 1989

\end{thebibliography}
\end{document}